\documentclass{emulateapj}

\usepackage{natbib}
\usepackage[section]{placeins}

\shorttitle{Rising flux tubes}
\shortauthors{Jouve \& Brun}

\begin{document}

\bibliographystyle{plainnat}

\title{3D non-linear evolution of a magnetic flux tube in a spherical shell: Influence of turbulent convection and associated mean flows}

\author{L. Jouve \altaffilmark{1}}
\altaffiltext{1}{DAMTP, Centre for Mathematical Sciences, University of
  Cambridge, Cambridge CB3 0WA, United Kingdom}
\affil{Laboratoire AIM, CEA/DSM-CNRS-Universit\'e 
Paris Diderot, IRFU/SAp, 91191 Gif sur Yvette, France}

\email{laurene.jouve@cea.fr}

\and
\author{A.S. Brun}
\affil{Laboratoire AIM, CEA/DSM-CNRS-Universit\'e 
Paris Diderot, IRFU/SAp, 91191 Gif sur Yvette, France}

\email{sacha.brun@cea.fr}

\begin{abstract}
We present the first 3D MHD study in spherical geometry of the non-linear dynamical
evolution of magnetic flux tubes in a turbulent rotating convection
zone. These numerical simulations use the
anelastic spherical harmonic (ASH) code. We seek to understand the
mechanism of emergence of strong toroidal fields through a turbulent
layer from the base of the solar convection zone to the surface as
active regions. To do so, we study numerically the rise of magnetic toroidal flux
  ropes from the base of a modelled convection zone up to the top of our
  computational domain where bipolar patches are formed. We compare the dynamical behaviour of flux tubes in a
fully convective shell possessing self-consistently generated mean
flows such as meridional circulation and differential rotation, with
reference calculations done in a quiet isentropic
zone. 

We find that two parameters influence the tubes
during their rise through the convection zone: the initial field
strength and amount of twist, thus confirming previous findings in Cartesian geometry. Further, when the tube is
sufficiently strong with respect to the equipartition field, it rises almost radially independently of the
initial latitude (either low or high). By contrast, weaker field cases indicate that downflows
and upflows control the rising velocity of particular regions of the
rope and could in principle favour the emergence of flux through
$\Omega$-loop structures. For these latter cases, we focus on the
orientation of bipolar patches and find that sufficiently arched structures are able to create bipolar regions with a predominantly East-West orientation. Meridional flow seems to determine
the trajectory of the magnetic rope when the field strength has been
significantly reduced near the top of the domain. Appearance of
  local magnetic field also feeds back on the horizontal flows thus
perturbing the meridional circulation via Maxwell stresses. Finally
differential rotation makes it more difficult for tubes introduced at
low latitudes to reach the top of the domain.
\end{abstract}


\section{Introduction}

At the solar surface, strong magnetic fields emerge during the whole
cycle, creating huge active regions with well-defined morphological
and dynamical characteristics revealed by high resolution observations
such as the 1m Swedish telescope in La Palma \citep{Scharmer} and the Hinode space telescope \citep{Kosugi}. In particular, according to Joy's law, most bipolar structures statistically show an East-West orientation, with a small tilt angle of a few degrees, increasing with the latitude of emergence (thus decreasing with the sunspots cycle).   
These active regions are believed to
take part in the global dynamo process operating in the Sun, and are the
results of the buoyant rise of the strong toroidal fields generated at
the base of the convection zone (CZ) in the tachocline of shear via
the so-called $\Omega$-effect \citep{Moffatt, Parker, Browning}. 

Active regions are thus thought to be the results of magnetic fields
emerging at the photosphere during the whole sunspot
cycle. Observations indeed indicate that magnetic flux continuously
emerges at the solar surface at all scales
\citep[see][]{vanDriel}. Although the emergence rate at small scales
strongly dominates over the emergence rate at large scale (which produces
active regions), the time scale of these large structures is much
longer and they are thus likely to take part in the process of a
global reconfiguration of magnetic fields in the chromosphere and the
corona. Violent events like CMEs are a good example of the role of
flux emergence at large scale since in most models of solar ejections,
an emerging flux system is supposed to be the triggering
mechanism. Moreover, observations have shown that a certain amount of
helicity of the magnetic structure could almost always be detected
\citep{Schmieder} even if it seems to be relatively small (according
to Chae \& Moon \cite{Chae}, a winding number of no more than 0.75 is
usually observed over a whole active region). This particular
ingredient is also thought to be responsible for the onset of some
violent events like CMEs through the kink instability \citep[e.g.][]{Torok, Fan3}.

Understanding the dynamical
properties of these magnetic structures requires to investigate the
rising mechanisms of strong toroidal structures through the turbulent
solar convection zone \citep[see review of][]{Fan}.
 Many models carried out since the 80's relied
on the assumption that toroidal flux is organised in the form of
discrete flux tubes which will rise cohesively from the base of the CZ
up to the solar surface \citep[see][however for a less idealised view of the topology of
buoyant flux structures]{Cattaneo}. The first emergence models used the "thin flux tube
approximation" \citep{Spruit1} in which the flux tube was
treated as a one-dimensional magnetic object moving in an idealised solar
convective envelope under the influence of magnetic buoyancy, tension,
aerodynamic drag and the Coriolis force. These models enabled to
demonstrate that the initial strength of magnetic field was an
important parameter in the evolution of the tube and that the active
regions tilts could be explained by the action of the Coriolis force
on the magnetic structure \citep{DSilva, Fan94, Caligari}. In the
framework of thin flux tube, Choudhuri \& Gilman \cite{Choudhuri}
studied the evolution of magnetic flux from the base of the CZ in a
rotating background and showed that the trajectory of emergence was
linked to the initial magnetic field strength. Another parameter then appeared to be
fundamental for the dynamical evolution of a flux tube: the twist of
the field lines. In the absence of twist, the tube splits into two
counter-rotating vortex tubes that move apart from one another horizontally and
eventually cease to rise. This behaviour was analysed by Sch{\"u}ssler
\cite{Schussler} and Longcope et al. \cite{Longcope} and then Emonet
\& Moreno-Insertis \cite{Emonet} showed that a threshold for the
amount of twist could be derived, that would ensure the coherence of
the tube during its rise. Three-dimensional simulations of
$\Omega$-loops however showed that this threshold is reduced by a
sufficiently arched magnetic structure. This curvature is indeed also able
to counteract vorticity generation due to the gravitational
torque applied to the flux tube \citep{Wissink, Abbett}. 

More sophisticated
multidimensional models \citep{Fan2} 
 in Cartesian geometry were then developed and extended to the upper
 part of the CZ and the transition to the solar atmosphere
 \citep[e.g.][]{Cheung,Archontis, Magara, Martinez}.
However, very few computations \citep{Cline, Dorch, Fan2} were performed to study the influence of convective
turbulent flows on the dynamical evolution of flux ropes inside the
CZ and none was done in spherical geometry. The assumption that turbulent flows may not have any influence on
the flux rise is only valid if the field strength is sufficiently
in superequipartition compared to the kinetic energy of the strongest
downflows and this argument is yet to be tested. 
Above all, no model has ever self-consistently studied the effects of 
convection, rotation, mean flows, curvature forces and 3D in the full 
MHD approach. We propose to do so in this paper, using the ASH code. Such computations
will allow us to assess for the first time the role of 
hoop stresses, Coriolis force, convective plumes, turbulence,
advection or shear by mean flows and sphericity 
on the tube evolution and on the subsequent emerging regions, along with the usual parameters such as field strength,
 twist of the field lines or magnetic diffusion.

The article is organised as follows. In Sect. 2, we present the
details of the simulation setup, including the equations solved, the
background hydrodynamical model and the initial magnetic
conditions. In Sect. 3, we summarise the results obtained in the
isentropic case, which will represent our reference case to which the
convective cases will be compared. In Sect. 4, 5, 6 and 7 the results of the computations in a
fully convective zone are presented, with a particular focus on the
structure of emerging bipolar regions and the influence of mean flows
on the magnetic field and finally in Sect. 8, we discuss
the results and interpret them in terms of dynamics of active regions
in the Sun.

\section{The model}
\subsection{Anelastic MHD equations}

The simulations described here were performed with the anelastic
spherical harmonic (ASH) code. ASH solves the three-dimensional
anelastic equations of motion in a rotating spherical shell using a
pseudo-spectral semi-implicit approach
\citep[e.g.][]{Clune,Miesch0,Brun2}. It uses a Large-eddy Simulation (LES) approach, with
parametrisation to account for subgrid-scale (SGS) motions. These equations
are fully nonlinear in velocity and magnetic fields and linearised in
thermodynamic variables with respect to a spherically symmetric mean
state to have density $\bar{\rho}$, pressure $\bar{P}$, temperature
$\bar{T}$, specific entropy $\bar{S}$. Perturbations are denoted as
$\rho$, $P$, $T$ and $S$. The equations being solved are

\begin{equation}
{\bf\nabla}\cdot(\bar{\rho}{\bf v})=0,
\end{equation}
 \begin{equation} {\bf\nabla}\cdot{\bf B}=0,
\end{equation}
\begin{eqnarray}
\bar{\rho}[\frac{\partial{\bf v}}{\partial t}+({\bf v}\cdot{\bf
    \nabla}){\bf v}&+&2\Omega_0\times {\bf v} ]=-{\bf \nabla} P
+\rho{\bf g}\\ \nonumber
&+&\frac{1}{4\pi}{\bf (\nabla \times B) \times B}
-{\bf \nabla \cdot \cal D}-[{\bf \nabla}\bar{P}-\bar{\rho}{\bf g}],
\label{eqNS}
\end{eqnarray}
\begin{eqnarray}
\bar{\rho}& &\bar{T}\frac{\partial S}{\partial t}+\bar{\rho}\bar{T}{\bf
  v}\cdot{\bf \nabla}(\bar{S}+S)={\bf
  \nabla}\cdot [\kappa_r\bar{\rho}c_p{\bf
    \nabla} (\bar{T}+T)\\ \nonumber
&+&\kappa_{0}\bar{\rho}\bar{T}{\bf
    \nabla}\bar{S}+\kappa \bar{\rho} \bar{T}{\bf \nabla} S ]+\frac{4\pi\eta}{c^2}{\bf j}^2 
+2\bar{\rho}\nu\left[e_{ij}e_{ij}-\frac{1}{3}({\bf \nabla \cdot
v)}^2\right],
\end{eqnarray}
\begin{equation}
\frac{\partial {\bf B}}{\partial t}={\bf \nabla \times} ({\bf v\times \bf B})-{\bf \nabla \times}(\eta {\bf \nabla \times B})
\end{equation} 

\noindent where ${\bf v}=(v_r,v_{\theta},v_{\phi})$ is the local
velocity in spherical coordinates in the frame rotating at a constant
angular velocity $\Omega_{0}$, ${\bf g}$ is the gravitational
acceleration, ${\bf B}=(B_r,B_{\theta},B_{\phi})$ is the magnetic
field, ${\bf j}=(c/4\pi)({\bf \nabla \times B})$ is the current
density, $c_p$ is the specific heat at constant pressure, $\kappa_r$
is the radiative diffusivity, $\eta$ is the effective magnetic
diffusivity and $\cal D$ is the viscous stress tensor. As stated
above, the ASH code uses a LES formulation where
$\nu$ and $\kappa$ are assumed to be an effective eddy viscosity and
eddy diffusivity, respectively, that represent unresolved
SGS processes, chosen to accommodate the
resolution. The thermal diffusion $\kappa_0$ acting on the mean
entropy gradient occupies a narrow region in the upper convection
zone. Its purpose is to transport heat through the outer surface where
radial convective motions vanish \citep{Gilman81, Wong94}. To complete the set of equations, we use the
linearised equation of state

\begin{equation}
\frac{\rho}{\bar{\rho}}=\frac{P}{\bar{P}}-\frac{T}{\bar{T}}=\frac{P}{\gamma\bar{P}}-\frac{S}{c_p}
\end{equation}

\noindent where $\gamma$ is the adiabatic exponent, and assume the ideal gas law

\begin{equation}
\bar{P}={\cal R} \bar{\rho}\bar{T}
\end{equation}

\noindent where $\cal R$ is the ideal gas constant, taking into
  account the mean molecular weight $\mu$ corresponding to a mixture
  composed roughly of 3/4 of Hydrogen and 1/4 of Helium per mass. The reference or
mean state (indicated by overbars) is derived from a one-dimensional
solar structure model and is regularly updated with the spherically
symmetric components of the thermodynamic fluctuations as the
simulation proceeds \citep{Brun2002}. It begins in hydrostatic balance so the bracketed
term on the right-hand side of Eq.\ref{eqNS} initially
vanishes. However, as the simulation evolves, turbulent and magnetic
pressures drive the reference state slightly away from strict hydrostatic balance.  

Finally, the boundary conditions for the velocity are impenetrable and
stress-free at the top and bottom of the shell. We impose a constant
entropy gradient top and bottom for the isentropic case and for the
fully convective case, a latitudinal entropy gradient is
imposed at the bottom, as in Miesch et al. \cite{Miesch}.
In all cases, we match the magnetic field to an external potential
magnetic field at the top and the bottom of the shell \citep{Brun2}.

\subsection{Introduction of a flux tube}
\label{sect_maginit}

To compute our model, we introduce at the starting time a torus of
magnetic field in entropy and total pressure equilibrium with the surrounding medium at
the base of the computational domain and we let the MHD simulation
evolve. 
We can derive an indication for the efficiency of the magnetic buoyancy in this situation of entropy and pressure equilibrium in writing the following relations respectively for the total pressure and the entropy equilibrium:

$$
\frac{P^{g}_{in}}{P^{g}_{ext}}=\frac{P^{g}_{ext}-P^{mag}}{P^{g}_{ext}}=1-\frac{B^2}{8\pi P^{g}_{ext}}
$$

$$
c_v \ln P^{g}_{in}-c_p \ln \rho_{in}=c_v \ln P^{g}_{ext}-c_p \ln \rho_{ext}
$$

These two equalities lead to the following relation between pressure and density inside and outside the flux tube:
$$
\frac{P^{g}_{in}}{P^{g}_{ext}}=(\frac{\rho_{in}}{\rho_{ext}})^{\gamma}
$$

\noindent with $c_p$ the specific heat at constant pressure, $c_v$ the specific heat at constant volume and $\gamma=c_p/c_v > 1$ the adiabatic index.

We can thus derive an expression for the density ratio between the
tube and its surroundings, as a function of the field strength:

$$
1-\frac{B^2}{8\pi P^{g}_{ext}}=(\frac{\rho_{in}}{\rho_{ext}})^{\gamma}
$$
$$
\frac{\rho_{in}}{\rho_{ext}}=(1-\frac{B^2}{8\pi P^{g}_{ext}})^{1/\gamma} 
$$
\noindent which gives a temperature ratio of:
$$
 \frac{T_{in}}{T_{ext}}=(1-\frac{B^2}{8\pi P^{g}_{ext}})^{\frac{\gamma-1}{\gamma}}
$$

We then note that under these conditions, the tube is introduced at a
slightly lower temperature than the surroundings, making it slightly
less buoyant than if it was introduced at pressure and temperature
equilibrium. For a tube introduced at $B=3.10^5 \, \rm G$ in a medium
where $P^{g}_{ext}=5. 10^{13} \, \rm dynes.cm^{-2}$ and
$T_{ext}=2.10^6 \rm K$ and with an adiabatic index of $\gamma=5/3$,
which are typical values for our simulations, the temperature
difference between inside and outside the flux tube is about
$40 \rm K$, thus significant with respect to the typical temperature fluctuations in our
convective flow.

\begin{figure*}
	\centering
	\includegraphics[height=7cm,angle=-90]{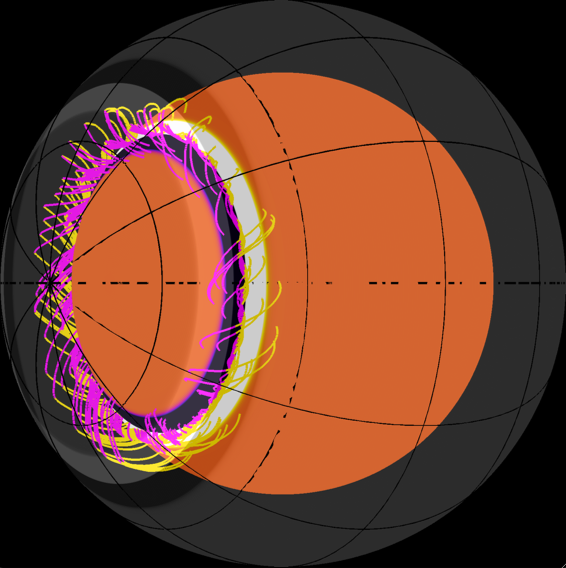} 
	\caption{Initial configuration of the magnetic flux tube
	  introduced at $45{\degr}$ and $R_t=5.2 \times 10^{10}\,
	\rm cm$ and with a twist parameter q=15 (corresponding to a
	pitch angle of $30{\degr}$). Purple indicates positive radial field and yellow indicates negative radial field.}
	\label{fig_init}
\end{figure*}

In this paper, we will not address how such coherent idealised
magnetic flux tubes are created within the Sun \citep[see][for details about magnetic buoyancy simulations and the creation of
buoyant arched structures]{Brummell02, Silvers09}. This regular
axisymmetric magnetic structure is embedded in an magnetised
stratified medium.  In order to keep a divergenceless magnetic field,
we use a toroidal-poloidal decomposition,

\begin{equation}
{\bf B}={\bf \nabla\times\nabla\times}(C {\bf e}_r) +{\bf
\nabla\times}(A {\bf e}_r)
\end{equation}

\noindent the expressions used for the potentials $A$ and $C$ for the flux tubes are:

\begin{equation}
A=-A_{0} \, r \, \exp\left[-\left(\frac{r-R_t}{a}\right)^2\right]
\times
\left[1+\tanh\left(2\frac{\theta-\theta_t}{a/R_t}\right)\right]
\end{equation}

\begin{equation}
C=-A_{0} \frac{a^2}{2} \, q \,
\exp\left[-\left(\frac{r-R_t}{a}\right)^2\right] \times
\left[1+\tanh\left(2\frac{\theta-\theta_t}{a/R_t}\right)\right]
\end{equation}

\noindent where $A_0$ is a measure of the initial field strength, $a$ is the tube radius,
 $(R_t,\theta_t)$ is the position of the tube center and $q$ is the
 twist parameter. The initial configuration of magnetic field is represented on Fig \ref{fig_init}.

Let's take for simplicity $\theta=\theta_t=45{\degr}$. Given the relations between the potentials $A$, $C$ and the three components of the magnetic field $B_r$, $B_\theta$, $B_\phi$, we can find an expression of the tangent of the pitch angle $\psi$ (angle between the
 direction of the vector magnetic field and the longitudinal
 direction) with respect to the initial parameters.
 
 $$
 B_r(r,\theta_t)=\frac{A_0 a q R_t}{r^2}\exp\left[-\left(\frac{r-R_t}{a}\right)^2\right]
 $$
 $$
 B_\theta(r,\theta_t)=\frac{2 A_0 q R_t (r-R_t)}{a r}\exp\left[-\left(\frac{r-R_t}{a}\right)^2\right]
 $$
 $$
 B_\phi(r,\theta_t)=\frac{2 A_0 R_t}{a}\exp\left[-\left(\frac{r-R_t}{a}\right)^2\right]
 $$
 
 \noindent Hence 
 
 $$
 \tan\psi=\frac{\sqrt{B_r^2+B_\theta^2}}{B_\phi}=q \frac{\sqrt{a^4+4(r-R_t)^2 r^2}}{2r^2}
 $$
 
The pitch angle is then linked to the parameter $q$ (appearing in the
expressions for the potentials $A$ and $C$) via a function of the tube
radius and position. Thus, we note that the pitch angle reaches its
maximum at the tube periphery (at $r=R_t+a$) and that it is close to
$0$ at the tube center (at $r=R_t$). We shall note at this point that
when $r=R_t+a$ (at the tube periphery), the term $a^4$ (due to the
contribution of $B_r$) becomes very weak in comparison to the other
term $4(r-R_t)^2 r^2$ (due to the contribution of $B_\theta$). The tangent of the maximal pitch angle is then approximately determined by the ratio $B_\theta/B_\phi$ and is in this case equal to $q a/(R_t+a)$.
 
We can now derive an expression for the winding degree of the field
lines (i.e. the number of turns that the field lines make over the
whole tube length $2\pi R_t \sin \theta_t$): 

$$
n=\frac{\pi R_t \sin\theta_t}{2 a}\tan\psi
$$

In all cases, except for Section \ref{sect_dif}, the tube radius is
set to $a=10^9 \,\, \rm cm$, about a twentieth of the depth of the
modelled convection zone and is introduced at the base of the CZ at
$R_t=5.2 \times 10^{10}\,\, \rm cm$. The initial field strength $A_0$,
the initial twist of the field lines $q$ as well as the colatitude of introduction $\theta_t$ will be varied in our models to investigate the influence of these various parameters.

\subsection{The background hydrodynamical models}
\label{sect_hydro}

Our experiments consist in introducing the torus of magnetic field at
the base of the convection zone in a spherical shell, as was presented above, in a thermally
equilibrated hydrodynamical model in which the convection is or is not
triggered. We then compute two different hydrodynamical models, one
which is isentropic and
one where we trigger the convection instability. The study of the
isentropic case is the topic of Jouve \& Brun \cite{Jouve} and will be considered as the reference case to
which the fully convective cases will be compared to.

Our numerical models are intended to be a faithful if highly
simplified descriptions of the solar convection zone. Solar values are
taken for the heat flux, rotation rate, mass and radius and a perfect
gas is assumed since the upper boundary of the shell lies below the H
and He ionisation zones. Contact is made with a real solar structure
model for the radial stratification. The computational domain extends
from about $0.72 R_{\odot}$ to $0.96 R_\odot$. The reference state was
obtained through the 1D CESAM stellar evolution code \citep{Morel97}
which uses a classical mixing-length treatment calibrated on solar
models to compute convection. We are dealing with the
central portion of the convection zone but neglect for this work the
penetrative convection below that zone or a stable top atmosphere. 

The effective viscosity and diffusivity $\nu$ and $\kappa$ are here taken to be functions of radius alone and
are chosen to scale as the inverse of $\bar{\rho}^{1/3}$. We use the
values: $\nu=1.13 \times 10^{12}\, \rm cm^2.s^{-1}$ and $\kappa=4.53
\times 10^{12}\, \rm cm^2.s^{-1}$ at mid-CZ, leading to a Prandtl
number of $P_r=0.25$. In all cases, the spherical shell is rotating at
the rate $\Omega_0=2.6 \times 10^{-6} \, \rm rad.s^{-1}$
(corresponding to a rotation period of 28 days).
In the convective cases, we trigger convection by assuming a Rayleigh
number $Ra=1.85 \times 10^5>Ra_c$ and setting a small and negative
$dS/dr=-10^{-7}$. In these cases, we have
$R_e=v_{conv}(r_{top}-r_{bot})/\nu_{midCZ}=120$, where the characteristic length scale is chosen to be the depth of
the CZ and $v_{conv}=80 \, \rm m.s^{-1}$. In the simulations, the Taylor number
is $T_a=1.8 \times
10^6$ and the convective Rossby number is then $R_{oc}=Ra/(T_a P_r)=0.63
< 1$, thus ensuring a prograde differential rotation \citep{Brun1}. The density
contrast in this convective case is about 24 whereas it reaches a
value of 40 in the isentropic case between the top and the bottom of
the domain.

\begin{figure}
	\centering
	\includegraphics[angle=90,width=7.7cm]{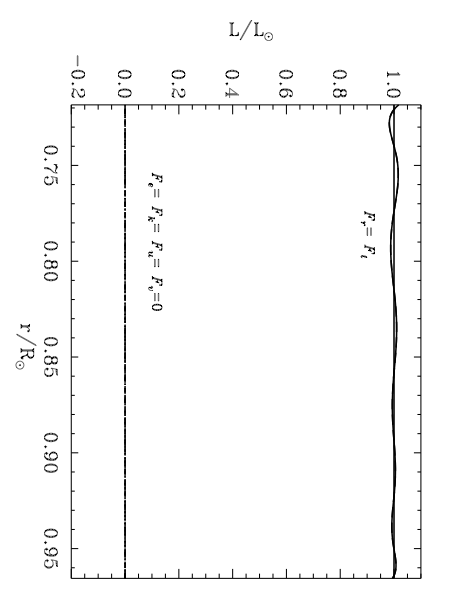}
	\includegraphics[width=7.5cm]{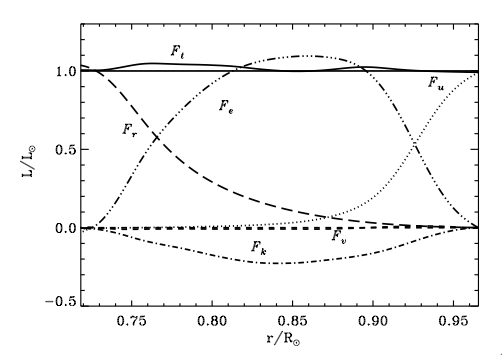}
	\caption{Radial dependences of the main fluxes involved in the
	isentropic (left) and the fully convective (right) cases.} 
 \label{figure_flux}
\end{figure}

Figure \ref{figure_flux} illustrates the contribution of various
physical processes to the total energy flux through the shell,
converted to luminosity and normalised to the solar luminosity, in
both the isentropic and the convective model. The
net luminosity, $L(r)$, and its components are defined as

\begin{equation}
F_e+F_k+F_r+F_u+F_{\nu}=F_t=\frac{L(r)}{4\pi r^2}
\end{equation}
\noindent where
\begin{equation}
F_e=\bar{\rho}c_p\overline{v_rT},
\end{equation}
\begin{equation}
F_k=\frac{1}{2}\bar{\rho}\overline{v^2v_r},
\end{equation}
\begin{equation}
F_r=-\kappa_r\bar{\rho}c_p\frac{\partial \bar{T}}{\partial r},
\end{equation}
\begin{equation}
F_u=-\kappa_0\bar{\rho}\bar{T}\frac{\partial \bar{S}}{\partial r},
\end{equation}
\begin{equation}
F_{\nu}=-\overline{{\bf v\centerdot\cal{D}}}\vert_{r},
\end{equation}

\noindent where
$F_e$ is the enthalpy flux, $F_k$ is the kinetic energy flux, $F_r$ is
the radiative flux, $F_u$ is the unresolved eddy flux, $F_{\nu}$ is
the viscous flux. 
The thermal diffusivity $\kappa_r$ is derived from a 1D
  calibrated solar
  structure model \citep{Brun2002} computed with the CESAM stellar evolution code \citep{Morel97}. We adjusted it so that the radiative flux is equal to
  the total flux in the whole layer in the isentropic case and is equal to the
  total flux at the base of the CZ for the convective case. In the
  latter case, the
  adjustment compared to the value obtained from the 1D model is small.
The unresolved eddy flux $F_u$ is the heat flux due to SGS motions,
which in our LES approach, takes the form of a thermal diffusion
operating on the mean entropy gradient \citep{Gilman81, Wong94}. Its main purpose is to transport energy outward through the impenetrable upper boundary where the convective fluxes $F_e$ and $F_k$ vanish and the remaining fluxes are small.

On Fig.\ref{figure_flux}, the flux balance is represented at the time
the system has reached a statistical steady state.  In the isentropic case,
as we do not have convection, we note that the energy is exclusively
transported by radiation, explaining why the total flux is equal to
the radiative flux in this case.

Contrary to the isentropic case, we note that several
fluxes play a role in the fully convective model. The
convective flux has developed to reach an equivalent luminosity of
almost 110$\%$ of the solar luminosity in the middle of the shell and
the radiative and unresolved eddy fluxes carry the energy at,
respectively, the bottom and the top of the domain where the enthalpy
flux vanishes. The viscous flux $F_{\nu}$ is relatively
small and slightly negative in most of the domain and the kinetic energy flux $F_k$ is, on the contrary, clearly negative in the whole convection zone. The asymmetry
between the fast downflows and the broad slower upflows is responsible
for the fact that the kinetic energy flux is negative. The very low
value of $F_{\nu}$ confirms that the Reynolds number of these simulations
is much greater than unity.

In the convective case, where different physical processes play a
significant role to transport the energy, large scale flows such as
differential rotation and meridional circulation (MC) are being
created due to the action of convective motions. In Brun \& Toomre (2002),
it has been shown that convection under the influence of rotation leads to an efficient redistribution
of angular momentum, energy and heat. It is found that Reynolds stresses are at the origin of the equatorial acceleration of
the solar convection zone, opposed by both the meridional circulation and viscous transport, the latter being negligible in 
the latest solar simulations \citep{Miesch08}. 
It was also found that the latitudinal enthalpy (convective) flux is at the origin of the variation
of entropy and temperature as a function of latitude, leading to warm
poles and cool equatorial regions \citep{Brun1, Brun08}. These thermal variations yield baroclinic effects that break the Taylor-Proudman
constraint of invariance along the axis of rotation.

\begin{figure*}
	\centering
	\includegraphics[angle=90,height=6.5cm]{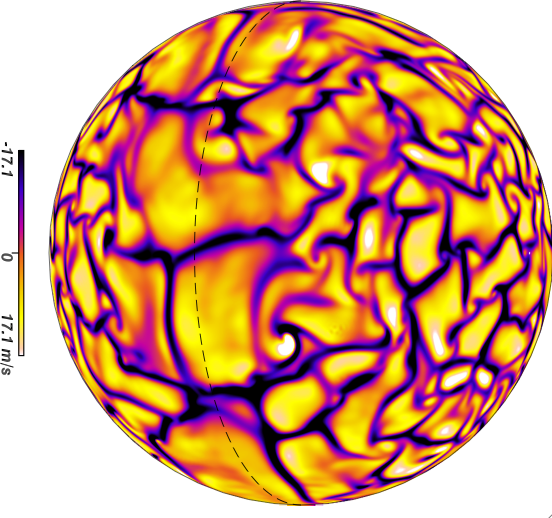}
	\includegraphics[angle=90,height=7cm]{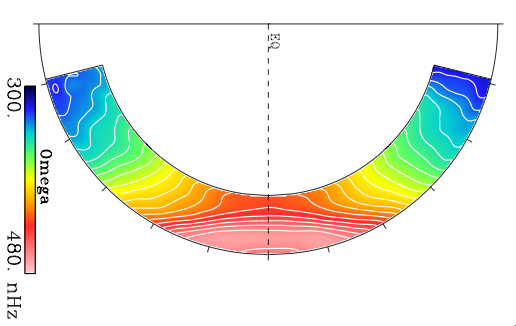}
	\includegraphics[height=7cm]{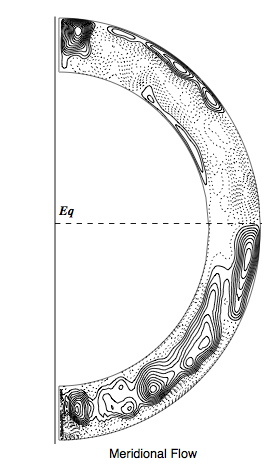} 
	\caption{Convective
	motions and mean flows created in the fully convective
	cases. The first panel shows the radial velocity profile near
	the top of the shell, the second shows the differential
	rotation profile and the right panel shows the meridional
	circulation, the last two panel having been averaged over
	longitude and time (272 days).
	For the meridional flow, dashed (plain) lines represent
	counterclokwise (clockwise) circulation and the intensity
	varies approximately between about -20 and 20 $\rm \, m.s^{-1}$.}
	\label{figure_lsf}
\end{figure*}

 Figure
\ref{figure_lsf} illustrates the convective structure and the
associated differential rotation and meridional flow realised in our
simulations. The convective patterns are complex, time dependent and
asymmetric owing to the density stratification, consisting of
relatively weak, broad upflows with narrow, fast downflows around
their periphery. By imposing in this model a weak entropy variation
at the base of the convection zone, which mimics the presence of the tachocline, we were able to get an even more 
 solar-like angular
velocity profile \citep[see Miesch][]{Miesch}. The relative
amplitude of this imposed variation corresponds to a pole-equator
temperature difference of about 10 K. The second panel of
Fig. \ref{figure_lsf} thus shows the differential rotation profile
which is in good agreement with the solar internal rotation profile
inferred from helioseismology \citep{Thompson}. In
this figure, the angular velocity of the rotating frame is 414 nHz,
corresponding to a rotation period of 28 days, the angular velocity
contours at mid-latitudes are nearly radial and the rotation rate
decreases monotonically with increasing latitude as in the Sun.  The
induced meridional circulation shown on the right panel of
Fig. \ref{figure_lsf} exhibits a complex profile, multicellular both
in latitude and in radius. Nevertheless, close to the equator, a
poleward flow of about $20 \, \rm m\,s^{-1}$ strongly dominates at the
surface, which is in agreement with helioseismic inversions.
The fully convective model is thus far more complex that the isentropic one. 
Continuously the dynamics is maintained with turbulent convection, 
heat and angular momentum redistribution, leading to the presence of
large-scale flows and asymmetric up and down flows whose various effects on a magnetic
flux rope will be studied.

\section{Evolution of a flux tube in an isentropic layer}
\label{sect_isen}

In this section, we briefly summarise the results obtained in the
calculations of Jouve \& Brun \cite{Jouve} concerning the influence of
the twist of the field lines, of rigid rotation and of the initial
latitude of the flux rope on its dynamical evolution in a stably
stratified layer. That will ease the comparison with the convective
case and complete our study with new isentropic models. Indeed, we
have decided that the simulations would be more realistic if the tube
radius was reduced to $10^9 \,\rm cm$. In section \ref{sect_dif}, we
will comment specifically on the effects of the tube radius on its evolution.

 Emonet \& Moreno-Insertis \cite{Emonet} showed that vorticity
 generation in the flux tube was controlled by the competition between
 the gravitational torque and the magnetic tension.  Consequently, by
 setting the gravitational torque to be equal to the projection of the
 Lorentz force in the equation for the azimuthal vorticity, we can
 determine the threshold above which the twist of the field lines can
 counteract the creation of two counter vortices inside the
 tube. We derive the following inequality for the pitch angle value:

\begin{equation}
\label{eqth}
\sin\psi=\frac{\sqrt{(B_r^2+B_\theta^2)}}{B} \geq \sqrt{\frac{a}{H_p}}\times\sqrt{\left|\frac{\Delta\rho}{\bar{\rho}}\right| \frac{\beta}{2}}=\sin\psi_{min}
\end{equation}
where $H_p$ is the pressure scale height at the base of the CZ,
$\Delta\rho/\bar{\rho}$ is the density deficit inside the tube compared to
the background stratification divided by the background density at the
tube center and $\beta$ is the plasma-$\beta$ associated with the
tube. In our case, the threshold value is equal to $0.3$
(corresponding to a pitch angle of 17.4$\degr$). In most twisted cases,
we use for $\sin\psi$ a value of $0.5$ (corresponding to a pitch
angle of 30 $\degr$), i.e. well above the threshold, so that the tube is
able to rise cohesively through the entire CZ.

Rotation has also an important dynamical effect on the
trajectory of the tube.  Indeed, as shown in Jouve \& Brun \cite{Jouve} in the non-rotating case, the
latitudinal component of the magnetic curvature acts to drag the tube
poleward as it cannot be compensated by any equatorward force (hoop stresses).  In the
rotating case, a retrograde zonal flow is created inside the tube
which induces a Coriolis force directed towards the Sun's rotation
axis which acts to deflect the trajectory of the tube poleward. Thus,
we note that the deviation to the radial trajectory in this case is
even more pronounced.

Moreover, we also showed in Jouve \& Brun \cite{Jouve} that the rotation has an impact on the rise time of
the tube.  Indeed, the radial component of the centrifugal force
decreases the tube velocity so that after 6 days of evolution, the
rise velocity of the tube in the non-rotating case is about 1.5 times
that of the tube in the rotating case.  This can be explained by the
fact that the buoyancy term is modified by an extra term coming from
the rotation which has the effect of limiting the efficiency of
buoyancy. The flux tubes thus emerge more slowly in the rotating
case.

\begin{figure*}
	\centering
	\includegraphics[width=14cm]{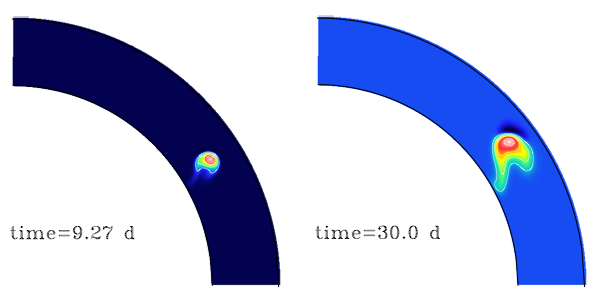}
	\caption{Rising trajectory of a flux tube introduced with
	$B_0=10^5 \,\rm G$. We note that the tube tends to rise parallel to the rotation axis.}
	\label{figure_radial}
\end{figure*}

We also investigated how flux tubes react when they are introduced
at various latitudes.  We find that the poleward drift due both to the
uncompensated magnetic curvature force (hoop stresses) and the Coriolis force varies
as a function of the latitude of introduction of the tube.  As
Moreno-Insertis et al. (1992) indicate, we can understand the poleward
drift in writing the equation for the $\theta$-component of the
velocity in the non-rotating case, neglecting the advection terms:
   
  \begin{equation} \centering \frac{\partial v_{\theta}}{\partial
  t}=-\frac{B_{\phi}^2}{4\pi r \bar{\rho}} \cot\theta \end{equation}
     
     This equation indicates that the acceleration in the
     $\theta$-direction is proportional to $\cot\theta$ which is a
     decreasing function of $\theta$ between $0$ and $\pi/2$. As
     $\theta$ is here the colatitude, the acceleration at higher
     latitudes is thus more rapidly active than at low latitudes and
     as a consequence, the poleward drift is much more visible for a
     flux tube originally located at high latitudes.

As Choudhuri \& Gilman \cite{Choudhuri} first demonstrated using the
thin flux tube approximation and as Fan
(2008) and Jouve \& Brun \cite{Jouve} confirm with 3D MHD simulations, the initial strength of the magnetic
field introduced at the base of the convection zone has a strong
influence on the rising trajectory of the flux rope. We thus computed
a new isentropic case where the initial field strength is $10^5 \,\rm G$ and found that in this case, as illustrated on Fig. \ref{figure_radial}, the tube is strongly deviated from the radial trajectory and tends to follow a path which is parallel to the rotation axis.

\begin{figure}
	\centering
	\includegraphics[angle=90,width=7.5cm]{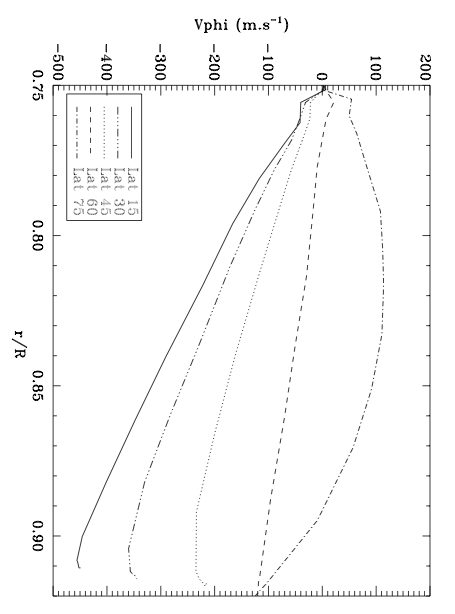}
	\caption{Intensity of the zonal flow in the axis of the flux
	tube with respect to its position, for tubes introduced at various latitudes.}
	\label{figure_vphi}
\end{figure}

The deviation to the radial trajectory is due to the creation inside
the tube of a retrograde zonal flow $v_\phi$ as soon as the magnetic
structure begins its rise through the isentropic layer, as illustrated
on Fig. \ref{figure_vphi}. The creation of this retrograde flow is the
result of the conservation of the total angular momentum $r \sin\theta
\bar{\rho} (r\sin\theta \Omega_0 + v_\phi)$ inside the tube. 
Its main effect is to locally create a Coriolis force oriented toward the solar rotation axis. This Coriolis force then partially compensate the component of the buoyancy force, perpendicular to the rotation axis whereas the component parallel to the rotation axis remains the same. As soon as this compensation becomes significant, the tube is strongly influenced by the component of the buoyancy force parallel to the rotation axis and thus drifts away from the radial trajectory.
We note that $v_{\phi}$ inside the tube is more and more negative to compensate for the creation of angular momentum due to the increase of $r\sin\theta$ for most cases. However, in the extremely high latitude case ($75{\degr}$), $v_{\phi}$ inside the tube first increases. This is due to the fact that for this case, the curvature force is first acting to make the tube drift poleward (since the curvature force acts faster at high latitudes), so that $\sin\theta$ decreases faster than $r$ increases, leading to a decrease of $r\sin\theta \Omega_0$ which has to be compensated by a prograde zonal flow. For this case, $v_{\phi}$ in the tube increases until $r\sin\theta$ becomes constant and then begins to increase and only then do we recover the same behaviour as tubes introduced at lower latitudes.
 
Nevertheless, the initial magnetic field strength plays a role in this
force balance. If the tube is weak like in the case of
Fig. \ref{figure_radial} where $B_0=10^5 \, \rm G$, the Coriolis force
created by angular momentum conservation is sufficiently strong to
compensate the weak buoyancy force and the main component which acts
on the tube will make it rise parallel to the rotation axis, which is consistent with the results of Fan (2008).

For the rise to be radial, we need to introduce sufficiently strong
magnetic tubes. We found a threshold of $1.3 \times 10^5 \,\rm G$ for the initial value of the magnetic field inside the tube. For the following calculations, we thus impose an initial value above this threshold so that active regions will emerge close to the latitude where the tube was introduced.

\section{Dynamical evolution of a flux tube in a fully convective shell}
\label{sect_conv}

\begin{deluxetable*}{cccccccccc}
\centering
\tablecolumns{10}
\tablewidth{0pc}
\tablecaption{Key parameters of the various convective cases}

\tablehead{
\colhead{Parameters} & \colhead{CAnt} & \colhead{CAtt} & \colhead{CAt} & \colhead{CBt} &
\colhead{CCt} & \colhead{CAt45} & \colhead{CAt15} & \colhead{CAt60} & \colhead{CAt75} } 

\startdata
$B_{0}$ & $5 B_{eq}$ & $5 B_{eq}$ & $5 B_{eq} $ & $10 B_{eq}$ & $2.5 B_{eq}$ & $5 B_{eq}$ & $5 B_{eq}$ & $5 B_{eq} $ & $5 B_{eq} $  \\ \hline 
$\Phi_0/10^{23}$ & $4.65$ & $4.65$ & $4.65$ & $9.45$ & $2.32$ & $4.65$ & $4.65$ & $4.65$ & $4.65$  \\ \hline
$Latitude$ & $60$ & $60$ & $60$ & $60$ & $60$ & $45$ & $75$ & $30$ & $15$   \\ \hline 
$\sin\psi$& $0$ & $0.33$ & $0.5$ & $0.5$ & $0.5$ & $0.5$ & $0.5$ & $0.5$ & $0.5$  \\ \hline 
$Re_t=\frac{v_{rise} a}{\nu_{midCZ}}$ & $16$ & $16$ & $16$ & $68$ & $4$ & $16$ & $15$ & $16$ & $17$  \\ 

\enddata
\label{tab}
\end{deluxetable*}

Our reference isentropic case has been defined and studied. We now
know that a sufficient twist of the field lines and magnetic field
intensity were necessary to enable the flux tube to rise cohesively
and radially through the isentropic layer. 
We now turn to investigate the evolution of similar tubes in a fully
convective zone where mean flows are developed and maintained.

\subsection{Description of the convective cases}

We compute a series of models
where the tube is introduced in the CZ after the convection and the
mean large scale flows have self-consistently developed and we compare
the results with reference cases in which we do not have convection.
We thus compute an untwisted case (the initial field is exclusively
oriented in the direction of the tube, i.e. $q$=0), a twisted case
(with a twist above the threshold of Eq. \ref{eqth}), cases with tubes
located at different latitudes and cases with various initial field
strength.  The various cases
and the parameters used are summarised in Table \ref{tab}.

As we said, the tube radius is set to $10^9 \, \rm cm$, about
0.18 times the pressure scale height at the base of the CZ. The
magnetic diffusivity at mid-CZ is set to the value of $1.13 \times 10^{12} \rm cm^2 \,
 s^{-1}$, leading to a magnetic Prandtl number of $1$, $\eta$ is made
to vary as $1/\bar{\rho}^{1/3}$ like the other effective eddy
diffusivities, leading to a value of $7.95 \times 10^{11} \rm cm^2 \,
 s^{-1}$ at the base of the CZ. The magnetic diffusivity is kept the same for 
all runs (except in Sect \ref{sect_dif}, where we investigate 
the influence of this parameter), we thus have the
same value of the diffusive time associated to the flux tube for all
cases which is $a^2/\eta_{baseCZ} = 14.5$ days. In the convective cases, we express the initial
magnetic field strength in terms of the intensity of the magnetic
field which is in equipartition with the kinetic energy of the
strongest downflows, this $B_{eq}$ is approximately equal to
$6.1 \times 10^4 \,\rm G$. The twist of the field lines is expressed in
terms of the sine of the pitch angle and can be compared to the
threshold value calculated according to Eq. \ref{eqth} in the
isentropic case. Abbett et al. (2000) showed that this
threshold may be reduced if we introduce a sufficient curvature in the
magnetic structure we initially set at the base of the CZ. As we study
here the evolution of initially axisymmetric flux tubes and not
$\Omega$-loops, the 2D-threshold value for the twist will be
used. Nevertheless, since modulation in longitude is created in
certain cases by the convective motions, we may also obtain a
significant curvature of the rope in our simulations and thus a lower
amount of twist would likely be sufficient to maintain the tube coherence
during its rise.

\subsection{Interaction with convection in the standard case}

In this section, we first focus on the influence of
the convective motions on the tube evolution in case CAt i.e. when it is introduced at
a latitude of $30{\degr}$, with a fixed initial twist of about $23$ turns and an
initial field strength of $3 \times 10^{5}\,\,\rm G$ (i.e. $5 \,
B_{\rm eq}$).

\begin{figure*}
	\centering \includegraphics[width=16cm]{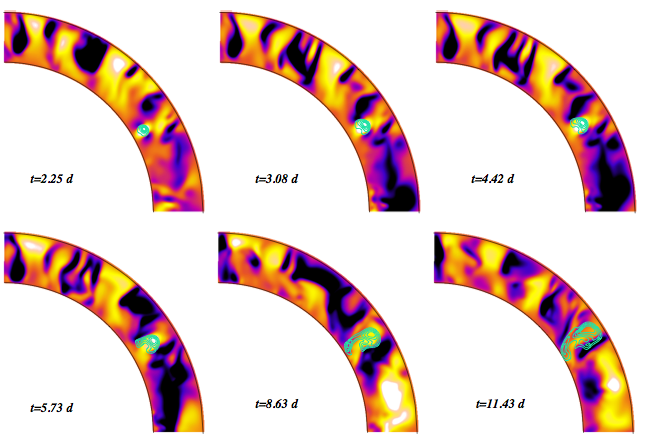}
	\caption{Evolution of $B_{\phi}$ cut at a specific longitude
	and shown in a portion of the Northern hemisphere, associated with the
	rising tube (contours) superimposed to the evolution of the
	background convection represented by the coloured contours of
	$v_r$. Blue (yellow) colours represent down (up) flows, the
	velocities vary from -300 $\rm \, m.s^{-1}$ to 200 $\rm \, m.s^{-1}$}.
	\label{figure_posl}
\end{figure*}

Figure
\ref{figure_posl} represents the contours of $B_{\phi}$ and of the
radial velocity $v_r$ as the tube rises through the CZ. We first
notice that the tube expands during its 12 days of evolution, to get to
a radial extension of about 3 times the initial one when the tube
reaches the top of domain. This expansion is due both to
magnetic diffusion and to the pressure drop from the base
to the top of the domain. This figure clearly shows the deformation of the
shape of the tube section while it rises. The magnetic initial
conditions, as we saw in Sect. \ref{sect_maginit}, imply a perfectly circular shape of the tube
section and after 12 days of evolution, the last panel of
Fig. \ref{figure_posl} shows that the tube has been squeezed at its
apex and thus develops an oblate shape during its rise. Moreover, the
periphery of the tube, where the magnetic field is much lower than in
the apex, has thus more difficulties to make its way through the
convective zone and consequently has the tendency to be dragged
downwards, contrary to the rising apex. If we look closely at the
convective pattern, the downward advection of the tube periphery can
be easily related to the downflows appearing at each side of the tube
as it rises. We then clearly see that the background convection is
strongly affected by the presence of the confined magnetic field. When
the magnetic structure begins its evolution, it creates its own local
velocity as we can see on the first panels of Fig. \ref{figure_posl},
due to the back reaction of the Lorentz force on the velocity
field. This velocity field due to the Lorentz force consists in a strong
upflow in the central region (the apex) and two downflows at each side
of the tube. The study of the momentum equation shows that this
particular configuration of the velocity is a direct consequence of
the presence of the latitudinal gradient of $B_r$ in the equation for
$v_r$, which changes sign twice across the tube section. In the
isentropic case, the same type of velocity field created by the
presence of the magnetic tube appeared during the evolution but in
this case, this configuration was much more symmetric with respect to
the apex of the tube since the only background velocity was due to the
tube. On the other hand, in the convective case, the velocity field
created by the tube is an additional velocity to the background
convection and thus a clear asymmetry is visible in the velocity field with
respect to the apex. Here, especially if we focus
on the 3 lower panels showing the last days of evolution, we note that
the two downflows at each side of the tube are very different in
extension and shape, leading to a very distorted aspect of the tube by
the time it reaches the top of the domain. The effect of the Coriolis
force consisting in deflecting the tube poleward is thus less visible
than in the isentropic case since now the velocity field in the
meridian plane also plays a very significant role in the dynamical
evolution of the flux tube, as we will discuss more in section
\ref{sect_meanflows}.

\begin{figure}
	\centering
	 \includegraphics[width=9cm]{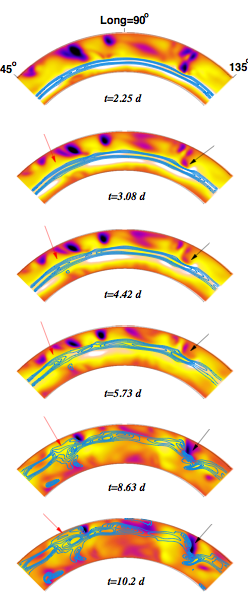}
	\caption{Evolution of the magnetic energy cut at the latitude
	of $30{\degr}$ (contours) superimposed to the
	background convection shown by a colour representation of
	$v_r$. Blue (yellow) colours represent down (up) flows, the
	velocities vary from -300 $\rm \, m.s^{-1}$ to 200 $\rm \, m.s^{-1}$. A black arrow indicates
	the location of a particular downflow a red arrow indicates the location of a particular
	upflow.}
	  \label{figure_eqsl}
\end{figure}

Since the background convection has a strong non-axisymmetric component,
it is likely that the evolution of the tube will depend on the longitude, contrary to the reference isentropic case. Indeed,
Fig. \ref{figure_eqsl} shows the evolution of the same flux tube with
the background convective motions, but projected on the $(r,\phi)$
plane. This view enables to observe the longitudinal deformation of
the magnetic field due to convective up and downflows. Since our aim
is to understand how active regions could be created at the solar
surface, the study of the
longitudinal deformation of the tube while it rises is of major
interest.

The cut of the magnetic energy and $v_r$ is made at $30{\degr}$ of latitude, where we
introduce the tube initially. We thus recover on each panel of
figure \ref{figure_eqsl} the strong upflow we could see on the
previous figure centered at the apex of the flux tube. Tracking a
particular upflow (red arrow on each panel) and a particular downflow
(black arrow) enables us to focus on the strong correlation existing
between the regions where the magnetic structure is lifted (pinned
down) and the convective upflows (downflows). Indeed, at the location
of the strong downflow, the field lines
are squeezed and thus retained in the solar interior, even if the tube
is still globally subject to magnetic buoyancy. We thus have a
competition between magnetic buoyancy and convective downflows which
controls the rising behaviour of the tube, as was seen in the
Cartesian study of Fan et al. (2003). In this region, even if the tube
locally creates an upflow, it is not sufficient to counteract the
strong background downflow and this portion of the structure is thus
clearly pinned down by convection and will eventually rise
significantly slower than the surrounding regions. On the contrary,
the strong upflow which owes its origin both to the background
convection and to the presence of the magnetic field clearly drags the
field lines upward and will most probably favour flux eruption at the photosphere. We moreover see
on Fig. \ref{figure_eqsl} that convective plumes
drift longitudinally in time due to the presence of rotation. This constitutes a major difference with previous 
Cartesian study where the convective structures were not influenced by rotation.
The same convective plume in our simulations will thus have effects on
the magnetic tube at different longitudes. A rough
calculation of the drifting time of convective plumes in the middle of
the convective zone at the latitude of $30\degr$ shows that a
convective plume could drift along $16\degr$ longitudinally during
this particular flux tube evolution (which lasts about 10 days). We
should however take into account that the presence of the flux tube
itself modifies the structure of the convective motions and thus may
influence the action of convective plumes on the magnetic field,
especially when it is introduced with higher intensity, as we discuss
in the following section.

\subsection{Interaction with convection when varying the field strength}

We have just seen that a modulation in longitude
appears as the tube rises through the turbulent layer, this modulation
is the result of strong interactions between convective motions and
the magnetic structure. These interactions are likely to be sensitive
to variations of the initial magnetic intensity inside the flux tube.
We thus investigate the influence of the initial magnetic field
strength in these fully convective cases. Few authors \citep[e.g.][]{Fan2,Murray} have already shown that
this parameter may have a strong influence on the rising behaviour of
the flux tube and on its interaction with the convective motions.

\begin{figure}
	\centering
	\includegraphics[width=7.cm,angle=90]{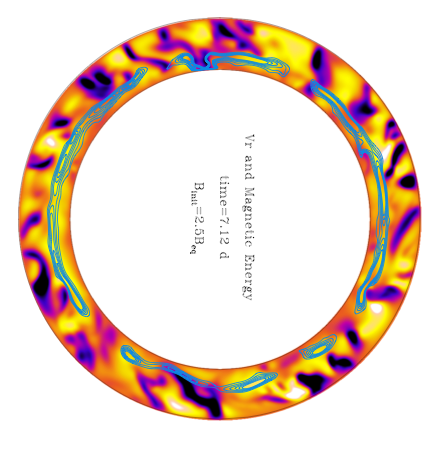}
	\includegraphics[width=7.cm,angle=90]{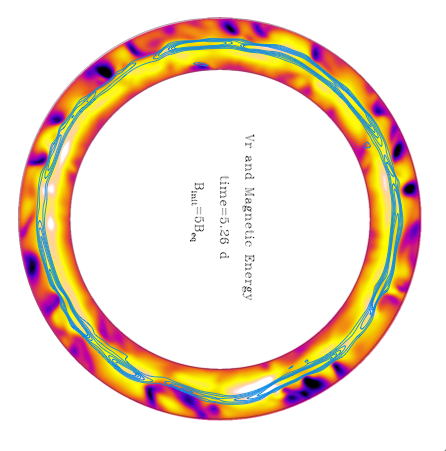}
	\includegraphics[width=7.cm,angle=90]{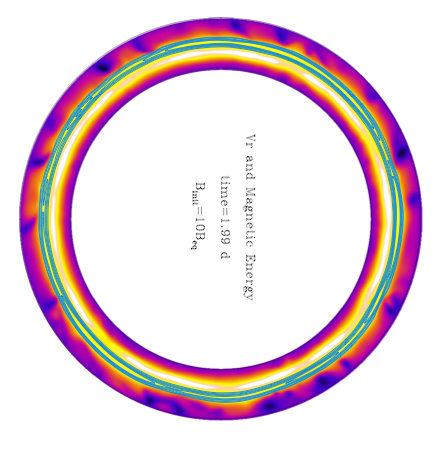} \caption{Cut
	at $\theta=45\degr$ of the radial velocity (colours) and of
	the magnetic energy (line contours) for 3 different
	initial values of the magnetic field strength (cases CCt, CAt, CBt). Yellow colours
	indicate upflows and blueish colours indicate downflows. The
	minimum velocity for all cases is about -300 $\rm \, m.s^{-1}$
	and the maximal velocity (concentrated inside the tube for the strong
	B cases) varies, its value is about 160 $\rm \, m.s^{-1}$ for
	case CCt, 200 $\rm \, m.s^{-1}$ for case CAt and 420 $\rm \,
	m.s^{-1}$ for case CBt.}
	\label{figure_updown}
\end{figure}

Figure \ref{figure_updown} shows the interaction between convective
motions and the rising behaviour of flux ropes introduced in the same
hydrodynamical background but with three different initial values for
the field strength. On the first panel, we show the result of the
calculation where $B_{init}$ is approximately equal to the
$2.5 B_{eq}$. In this case, the correlation between the upflows
(downflows) and the portions of the tube which rise more rapidly
(slowly) is clearly visible. The background velocity dominates over
the velocity field created by the flux rope through the Lorentz
force. As a result, it is the background velocity which controls the
rising behaviour of the tube. Since the initial field strength is relatively
weak in this case, the convective motions first deform the tube in
longitude, then the strong downdrafts pin the tube down and finally
the rope loses its buoyancy by magnetic dissipation before it is able
to rise through the entire convection zone. The rope is thus unable to
rise to the top of the domain in this case where B is $2.5$ times the equipartition field.  On the
contrary, when the field is very strong compared to the equipartition field 
(last panel of Fig.\ref{figure_updown}), the
background velocity field has almost no effect on the behaviour of the
rope. Its self-created velocity completely dominates the evolution and
thus the tube rises almost axisymmetrically as if it was embedded in a
stably stratified zone, even if a weak modulation in longitude is
visible on Fig \ref{figure_updown}.  In the intermediate case, where the field
strength is close to 5 times the equipartition field, we note that
the flux rope is strongly modulated in longitude but the whole tube
emerges anyway in a relatively coherent manner. In this case, the
velocity created by the tube itself is of the same order as the
background velocity and the convective motions are thus able to
strongly influence the tube during its dynamical evolution inside the
CZ. This is an interesting behaviour since even if the tube is
introduced axisymmetrically, some longitudes can be favoured and
structures will be able to emerge only in few places at the solar
surface, thus creating localised active regions.

As we saw, while it rises, the flux tube creates its own local
velocity field which may strongly disturb the background velocity
field, especially when the initial magnetic field intensity is strong
compared to that of the equipartition field. This explains why the
tube is more or less influenced by the convective motions as it
evolves in the CZ. Indeed, if the magnetic energy of the tube is
strong compared to the kinetic energy of the strongest downdrafts, the
tube creates a velocity through the action of the Lorentz force which
dominates against the background purely hydrodynamically-generated
velocity. Since a strong upflow is thus created near the tube axis,
the rising mechanism is very efficient and the tube reaches the top of
the CZ in only 4 days. In the weaker cases, the velocity field
created by the magnetic structure is comparable to the background
velocity field and the latter is thus able to influence the behaviour
of the flux rope as it rises, the rise time is in this case of about
12 days.

\section{Structure of the emerging regions}

We now turn to discuss the characteristics of active regions created
by the buoyantly rising magnetic structures. We especially focus on
the field strength, the orientation and the later evolution of the
bipolar active regions. However, it has to be clarified that
  since our upper boundary lies at about 28 Mm below the actual solar
  surface, what we call ``flux emergence'' here is the emergence
  through the top of the computational domain, which is likely to be
  different from the emergence in the real photosphere.

\subsection{Creation of bipolar regions in the standard case}

\begin{figure*}
	\centering 
	\includegraphics[width=15cm]{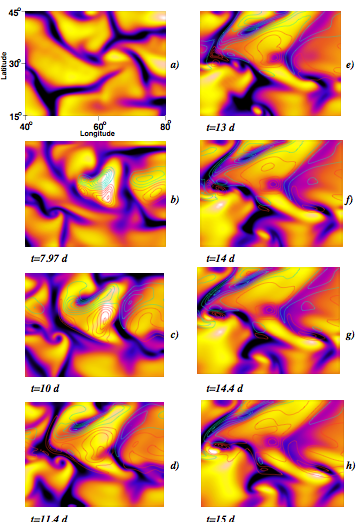}
	\caption{Evolution of the radial field (contour lines)
	  superimposed 
	  to the radial velocity (colour contours) at 0.96R and zoomed
	  in a
	  particular portion of the Northern hemisphere where 
	  a bipolar region emerges. Green (red) corresponds to
	  positive (negative) $B_r$. 
	  The first panel shows the field-free region prior to the
	  emergence 
	  and the other panels show the evolution of the magnetic
	  structure and 
	  the velocity field from the time of emergence (2nd panel) 
	  until approximately 7 days later (last panel).}
	\label{figure_shsl}
\end{figure*}

Figure \ref{figure_shsl} shows a zoom, seen from above, of an emerging bipole near the top of the domain. On this figure the radial field $B_r$ is shown, superimposed to the background radial velocity. We can here focus on the change in the convective patterns as the flux tube emerges, on the influence of particular downflows on the magnetic structure and on the late evolution of the magnetic field after the emergence.

On this figure, we clearly note that the emerging phase is
characterised by the appearance of the bipolar patch in a very
localised portion of the ($\theta,\phi$) plane which in turn locally modifies
the convective patterns. We again recover the strong upflow created by
the tube and located at its apex and the downflows which appear at
each side of the emerging tube. We then note that the convection
organises itself very differently around the magnetic field. The
strong central upflow significantly influences the background velocity
field: for example, the strong downflow located on panel a) (before
emergence) around the longitude of $75{\degr}$ and the latitude of
$30{\degr}$ is modified by the appearance of magnetic structures on panel
b), the downflow is squeezed and becomes less intense in the area where
the tube emerges. However, this downflow is so strong at the beginning
that in spite of the influence of the magnetic field, we recover its
imprint during the whole evolution on all the panels. On the other hand,
the upflows which were already present before the arrival at the top
of the domain of the magnetic structure are enhanced by the flux
emergence and for example the patch of positive radial velocity
located in the middle of the first panel stays very strong during the
whole evolution because it is reinforced by the emergence of the
bipolar structure. The magnetic field has thus a strong influence on
the modification of the convective patterns at the top of the domain
but we can also focus on the influence of convection on the
deformation of the rope and on its late evolution. Indeed, the strong
downflows are clearly the areas where the magnetic structures have
more difficulties to emerge whereas strong upflows make it very easy
for the bipolar patches to appear from the beginning. On panels
b) and c) in particular, we clearly see the emergence of two active
regions separated by a strong downflow which causes the central area
to stay deeper down in the interior. The tube thus emerges in a
particular shape, with 2 distinct bipolar patches appearing and then evolving differently.

The late evolution of the flux tube after emergence also presents some
interesting properties. On the first panels (b), c) and d)), the field
which is brought to the surface by magnetic buoyancy dominates the
evolution of the simulation since it strongly perturbs the convective
motions. At this stage, we can note that the tube evolution will
remain dominated by the convective motions since the diffusive time
scale at the surface for regions occupying about $10\degr$ in
longitude is of the order of 45 days, much higher than the advection
time (of the order of a few days).  
As the simulation evolves, the magnetic field begins to be
advected horizontally by convection which tends to separate the two
opposite polarities of the bipolar patches (panels e) and f)). Panels g)
and h) then show the behaviour of the magnetic field and the radial
velocity about 7 days after the first signs of emergence. On these
panels, the field lines become stretched by the convective motions and
advected towards the strong downdrafts. We then recover some features
of magneto-convection when the magnetic field is less organised than
one well-defined flux tube. Indeed, at this stage of evolution, the
structure is much less coherent, it begins to occupy a very wide band
in latitude because of the redistribution by convection.

\begin{figure*}
	\centering 
	\includegraphics[width=17cm]{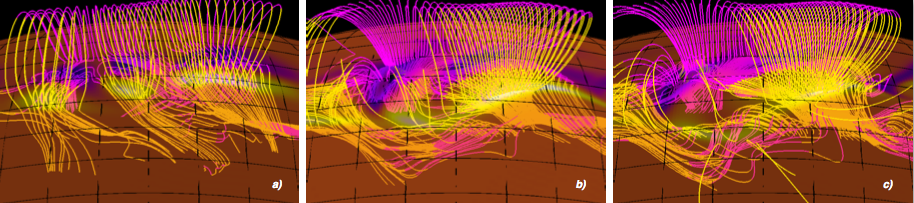}
	\caption{Magnetic field line reconstruction below the emerging
	regions and above, where a potential extrapolation has been
	applied. Yellow (purple) lines indicate positive (negative)
	radial field. The three snapshots correspond to approximately
	panels b, e and h of Fig. \ref{figure_shsl} }
	\label{figure_shsl3D}
\end{figure*}

Figure \ref{figure_shsl3D} enables to see the 3D emergence of the
bipolar structures, in showing the magnetic field lines reconstruction
immediately below and above the top of our computational
domain, at different times in the emergence process. Panel a)
corresponds to the first signs of emergence, we clearly note (like panel
b of Fig. \ref{figure_shsl}) the North-South orientation of the
bipolar patches, which becomes more and more East-West as the emergence
proceeds, as is shown on panels b) and c). On these panels, the
complicated structure of the flux rope starts to be visible in the
interior. Indeed, we note the modulation both in latitude and in
longitude of the tube when it reaches the top of the domain. Panel c)
shows that the magnetic field of the tube connects with the external
field during the emergence, even if some parts of the rope stay hidden in the
solar interior and are not able to rise anymore. In particular, the fact that the tube axis dos not
emerge and that only the upper part of the rope is visible outside the
computational domain is probably responsible for the predominantly North-South
orientation of the emerging patches. 

This remark thus leads us to
analyse in more details the evolution of the tilt angle and of other
characteristics of the emerging regions, in the same spirit of the
observational studies of Kosovichev \& Stenflo \citep{Koso2008}.

\begin{figure}
	\centering 
	\includegraphics[width=9cm]{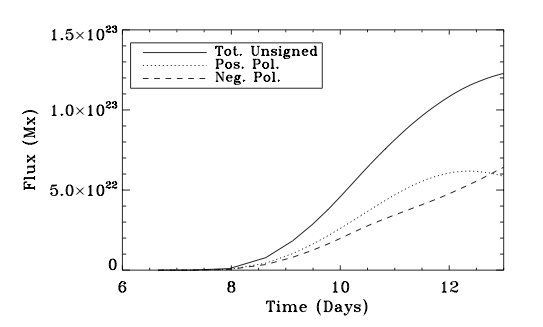}
	\includegraphics[width=8cm]{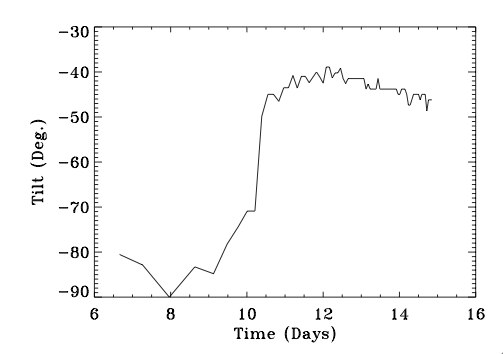}
	\includegraphics[angle=90,width=8cm]{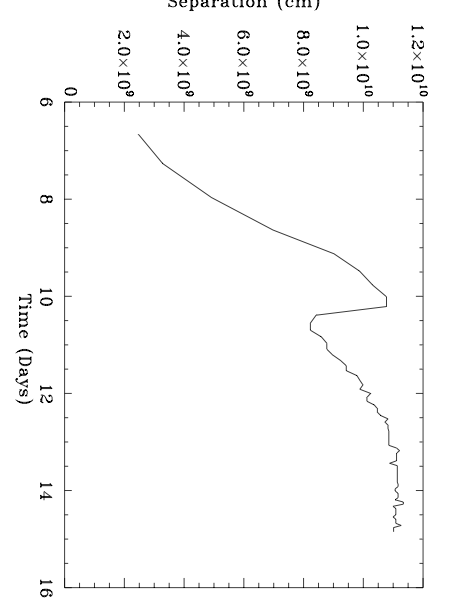}
	\caption{Magnetic fluxes, tilt angle and footpoints 
	  separation for one of the emerging regions}
	\label{figure_shslflux}
\end{figure}

Using a series of 96-min cadence magnetograms form SOHO/MDI,
they analysed 715 active regions in
terms of the evolution of the tilt angle, of the amount of emerging flux
and of the magnetic polarities separation during emergence. We
thus proceed to the same kind of analysis on our particular portion of
magnetic field emerging between the longitudes of 55 and 65 degrees on
panel b) of Fig. \ref{figure_shsl}. We have to keep in mind that the
  emergence through our upper boundary (which still lies well inside
  the CZ) is difficult to compare directly to observed flux emergence at the photosphere. However, this type of analysis
  enables us to get a better insight into the processes playing a role
  in the evolution of magnetic flux ropes well below
  the photosphere, thus allowing to predict some characteristics of the
  structures which will actually emerge through the upper layers. The results of this
analysis are shown on Fig. \ref{figure_shslflux}. This figure shows
the evolution of the total unsigned flux (together with the
contribution of the positive and negative polarities), the tilt angle
and the magnetic polarities separation. The amount of flux first
sharply increases during the first 3 or 4 days after the first signs
of emergence and then reaches a saturation and starts to decrease as the opposite magnetic polarities stop separating. After about 4 days after emergence, the separation between the two opposite polarities is not modified by emergence anymore and the concentrations of radial field start to be advected by convection and by magnetic diffusion on a longer time scale than the rise time of our flux rope, leading to a saturation of the footpoints separation visible on the last panel of Fig. \ref{figure_shslflux}. Finally, we investigate the evolution of the tilt angle of our emerging bipolar region and note that the orientation is mainly
North-South on the first days of emergence (the tilt angle is then
equal to about $-90\degr$). 
Bipolar regions are thought to be the imprints of the
flux tube axis emerging, creating a positive radial field at one foot
of the emerging loop and a negative radial field at the other foot. In
these simulations, we do not clearly see the axis of the tube
emerging, the radial field which is observed is the one existing at the apex of
the tube because it is twisted. However, as the emergence proceeds, we
see that the orientation of the bipolar structure changes because of
the convective motions, the 2 polarities are advected more and more
independently and the orientation becomes more East-West (a tilt angle
of $-40{\degr}$ is reached when the active region begins its decay)
both because the structure is made sufficiently arched by the radial
velocity and because the horizontal velocity acts differently on the 2
regions of opposite polarity. By applying the same kind of analysis
for particular active regions created by tubes initially located at
the latitudes of $45\degr$ (case CAt45) and
$15\degr$ (case CAt15), we can assess how the tilt angle changes as a
function of the initial latitude.  We do not see a clear difference in the tilt angle for case CAt45 in comparison to case CAt described above,
mainly because their rise time, the amount of flux emerging and the
arching of the magnetic structure are similar. We thus not clearly see
different effects of the physical processes involved to
modify the tilt angle (Coriolis force, advection by convection, twist
of the field lines) between these 2 cases. On the other hand, in case
CAt15, where the tube is introduced at the latitude of $15\degr$, the
tilt angle reaches a smaller value (about $-20\degr$ when the active
region starts to decay). This can be explained mainly by the
difference in the rise time of this particular tube compared to the
others, as we will discuss in detail in Sect. \ref{sect_difrot}. Since
the rise time for this tube is longer, the Coriolis force may have the
time to significantly affect the tube orientation when it reaches the
surface. Moreover as we will see in Sect. \ref{sect_dif}, magnetic diffusion plays the role of untwisting
the flux ropes. Since the rise time is longer in this case, diffusion
has acted more on this tube and the tube appears less twisted when it
emerges, which leads to a tilt angle reduced in comparison to the
cases at higher latitude. This particular feature may be in agreement with
Joy's law which states that bipolar structures emerging at lower latitudes
have a smaller tilt angle than regions emerging at higher latitudes at
the beginning of a new solar cycle.

\subsection{Influence of the field strength on emergence}

The initial magnetic field strength has a strong influence on the
way the structure will emerge as it reaches the top of the
  computational domain, thus
creating active regions with various morphological and dynamical
characteristics.

\begin{figure*}
	\centering \includegraphics[width=16cm]{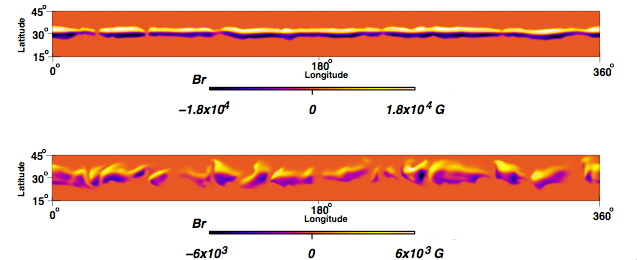}
	\caption{Cut at $r=0.93 R$ of $B_r$ for case CBt (lower panel) 
	  and case CAt (upper panel). We clearly note the
	difference between the two computations in the way the
	magnetic flux emerges.}  \label{figure_emerge}
\end{figure*}

Figure \ref{figure_emerge} shows the radial magnetic field close to
the top of the shell by the time the axis of the flux rope is situated
approximately at $0.93R$, for cases CAt and CBt. When
the tube is strong, the tube emerges at all longitudes with very small
azimuthal modulation even if the strong downflows have been able in
some portions of the tube to keep it from emerging as fast as in the
upflow regions. We also notice that the flux rope emerges at
approximately the latitude of introduction, no poleward slip is thus
visible in this case. On the contrary, in the weak B case, some
longitudes are clearly favoured and some 'active regions' can be
identified. 
We indicated on Fig. \ref{figure_emerge} the intensity of the emerging
$B_r$, which is about a few kiloGauss in case CAt. On the
  other hand, when the tube is introduced with a flux of about
$10^{24} \,\rm Mx$, as in the strong B case, the emerging radial field
is of the order of a few tens of kiloGauss. In this case, a
  strong flux loss would then have to be experienced by the tube
  during its rise up to the photosphere to match the observations of
  sunspots magnetic field at the solar surface.


We also note that in the weak B case, the latitude of emergence is
slightly higher than the latitude of introduction of the flux
tube at certain longitudes. Indeed, we see on the bottom panel of Fig.\ref{figure_emerge}
that the emerging structures appear at latitudes higher than the
latitude of introduction ($30{\degr}$). This drift could be explained partly by
the poleward slip instability and the action of the Coriolis force
which were already observed in the isentropic case but may also be due to
the action of the mean meridional flow as we now discuss in section \ref{sect_meanflows}.


\section{Influence of mean flows}
\label{sect_meanflows}

As stated in sect. \ref{sect_hydro}, convection in a spherical shell
establishes and continuously 
maintains mean flows. We wish to take benefit of our self-consistent
simulations to address the question of how meridional flows and
differential rotation may influence the tube-like structure during its
rise through the CZ.

\subsection{Differential rotation}
\label{sect_difrot}

The results coming from the study of flux tubes in a stably stratified
spherical layer in Jouve \& Brun \cite{Jouve} showed that the dynamical evolution of flux tubes could
be modified if they were introduced at various
latitudes. Indeed, for instance, we saw that the poleward drift was
more rapidly active for tubes introduced at high latitudes, thus
leading to a strong deviation of these tubes to their radial
trajectory. In the fully convective cases, we saw that the convective
patterns as well as the profile of the large-scale flows strongly
varied in latitude and longitude (see Fig.\ref{figure_lsf}). It is thus likely that the differences between
tubes introduced at various latitudes will be even more pronounced in
these cases.

\begin{figure}
	\centering \includegraphics[width=8.cm]{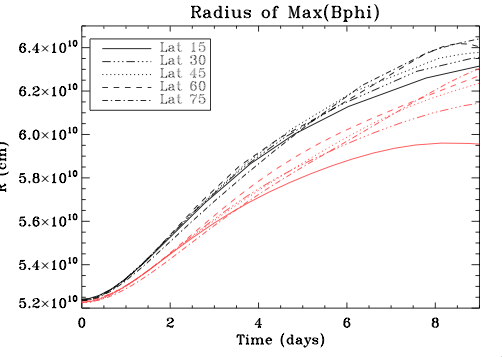}
	\includegraphics[width=8.cm]{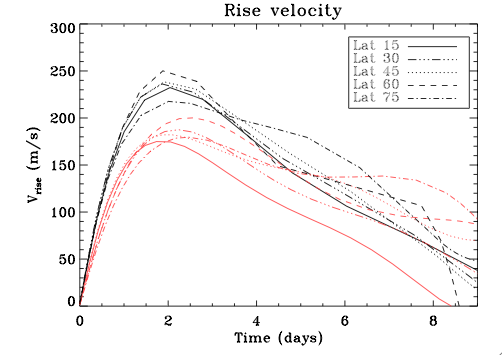}
	\includegraphics[width=8.cm]{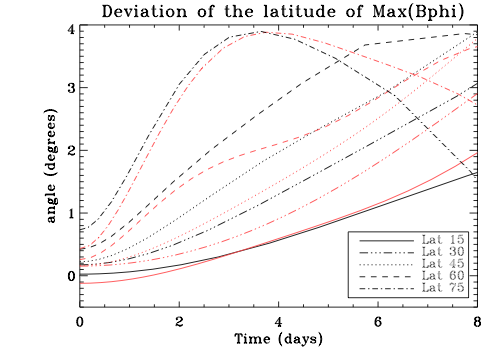}
	\caption{Comparison of the evolution of flux tubes introduced
	at different latitudes (cases CAt15, CAt, CAt45, CAt60, CAt75)
	in red with their isentropic counterparts in black.
 The first panel shows the position in
	radius of the maximum of $B_{\phi}$ versus time, the second
	panel is the velocity of the tube versus time and the last
	panel shows the difference between the position in latitude of
	the max of $B_{\phi}$ and the latitude of introduction.}
	\label{figure_latitude}
\end{figure}

Figure \ref{figure_latitude} shows the temporal evolution of weak
tubes initially located at the latitude of $15\degr$, $30\degr$,
$45\degr$, $60\degr$ and $75\degr$. The first panel shows the position
in radius of the maximum of $B_{\phi}$ (corresponding to the location
of the tube axis) as the tube rises through the CZ, the second panel
indicates the rise velocity of each tube and finally the third panel
represents the deviation in latitude of the position of $\rm
max(B_{\phi})$ compared to the latitude of introduction.  We confirm
 that tubes introduced at various latitudes
have different evolutions. Looking at panels 1 and 2, we
see that the tube initially located close to the equator (at $15\degr$) can be clearly distinguished from
the others, especially in the convective case. This tube indeed reaches its maximal velocity before the
others and the decelerating phase is so significant that it almost stops rising
 after it has reached the middle of the convection
zone. After 8 days of evolution, its rise velocity indeed becomes
very weak (about $10 \rm \, m.s^{-1}$) and the radial position of the
tube axis reaches its maximum at about $5.95 \times 10^{10} \, \rm cm$
in the CZ. On the other hand, the three tubes initially located in the
upper part of the Northern hemisphere keep on rising until the tube
periphery reaches the top boundary condition where the radial velocity
vanishes. This difference between tubes introduced at various latitudes is less significant in the isentropic case. For example, after 9 days of evolution in an isentropic layer, the distance travelled by a tube introduced at $15{\degr}$ is $5\%$ less than that of a tube introduced at $45{\degr}$. In the convective case, this difference reaches $20\%$.
This may be explained by the presence of a differential rotation in the convective case.
 We showed in the isentropic case that the buoyancy term
appearing in the evolution of the radial velocity was proportional to
$g-r\sin^2\theta \Omega^2$ with $\theta$ the colatitude and $\Omega$
the rotation rate. We showed that at constant $\Omega$, an increase in
the colatitude $\theta$ caused a decrease of this term and thus of the
efficiency of buoyancy, resulting in a slower emergence at higher
colatitude or lower latitude. In this convective case here, a
solar-like differential rotation is present in the bulk of the CZ. The
profile of this differential rotation is conical between $25\degr$ and
$60\degr$ and cylindrical under $25\degr$. This differential rotation
may explain the major reduction of velocity in the cases at low
latitudes in comparison to the cases at high latitudes as the strong
rotation at low latitudes is very likely to decrease the rise velocity
of the tube to such a point that it is not able to rise through the
upper part of the CZ. However, the effect of the centrifugal force
which modifies the buoyancy may be weak compared to the total
gravitational acceleration and thus the slowdown of tubes introduced
in a convective background could also be caused by the Coriolis force
due to the retrograde flow created along the tube.

Panel 2 shows that because of the convective downdrafts acting to pin
the flux tube down, the rising velocity is reduced compared to the
isentropic case. Indeed, when the tube was introduced at $45\degr$, it
reached a maximal velocity of about $230 \, \rm m.s^{-1}$ while the maximal 
velocity is only $180 \, \rm m.s^{-1}$ in the convective case,
i.e. $21\%$ less. 
For a tube introduced at the latitude of $15{\degr}$, the difference is even more pronounced and reaches the value of $27\%$, 
which in turn leads the tube embedded in the convective background to be stopped by the time it reaches the middle of the CZ.
The effects of intense downflows again appear to be very significant in
these weak B cases and we thus show the dramatic changes that the introduction
of a convective environment implies on the simulations of rising flux
tube. It has to be noticed that the ratio between the rise velocity of a section of the
tube located in an upflow and another located in a downflow is about
1.5. Thus for example in case CAt45 where the average rise velocity is
about $180 \, \rm m.s^{-1}$, the region of the tube located in a
particular upflow can reach a velocity of $210 \, \rm m.s^{-1}$
whereas a part located in a downflow will hardly reach $140 \, \rm
m.s^{-1}$. We note that these values are still lower than the rise
velocity reached by a tube introduced in a stable isentropic layer,
possibly showing the influence of the strong downflows over the whole
tube which tries to keep its coherent structure during its rise.
Our study clearly shows the effects of a non-uniform rotation on magnetic ropes, especially the 
severe constraints on low-latitudes emergence it introduces.
Although the emergence through our upper boundary may have little
  resemblance with emergence at the real solar surface, this
  particular finding may still be interesting to consider when
  analysing the properties of the solar cycle which shows a strong decrease
in the number of sunspots appearing at low latitudes in the declining phase. Only sufficiently
strong flux tubes would be able to rise at low latitudes, which is
confirmed by some observations 
of sunspot magnetic field during a cycle. Indeed, sunspots emerging at higher latitudes seem 
to possess brighter umbrae, thus indicating weaker magnetic fields
\citep{Norton04}, although this tendency seems to be slight and thus
possibly due to an observational bias \citep{Livingston} and has not
been confirmed by other space-based studies \citep{Mathew}.
If we suppose that such effects happening deep inside the convection
zone are visible during emergence at the surface, a possible explanation would thus be that differential rotation makes it more difficult for weak tubes to 
emerge at low latitudes and not only that they are drifting away from the radial trajectory as they rise,
as they conclude in Norton \& Gilman (2004).

\begin{figure*}
	\centering
	\includegraphics[width=16cm]{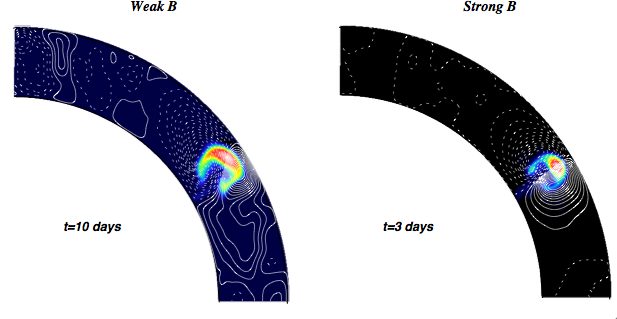}
	\caption{Cut of $B_\phi$ at a particular longitude in the
	  Northern hemisphere for cases CAt 
	  (left panel) and CBt (right panel) superimposed to the
	  background meridional flow. For the meridional flow, dashed (plain) lines represent
	counterclockwise (clockwise) circulation. We note the strong meridional
	  circulation 
	  created by the tube itself when B is initially strong.} 
\label{figure_circmerid}
\end{figure*}

\begin{figure*}
\centering
\includegraphics[width=18cm]{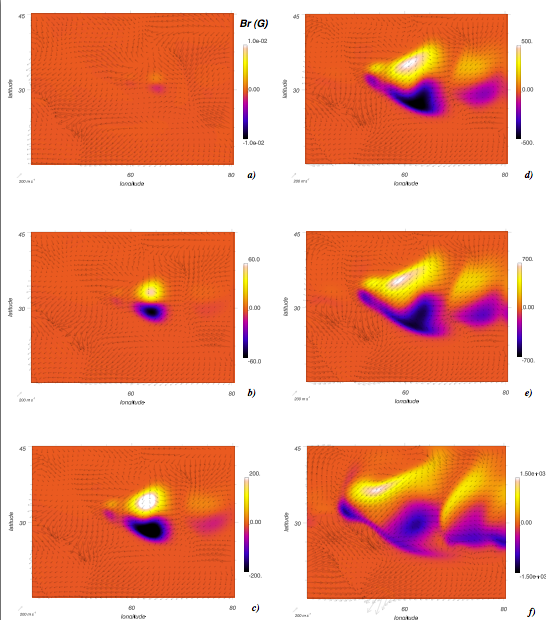}
\caption{Influence of the emerging magnetic flux on the surface flow
  structure. 
  We superimpose the radial field (coloured contours) 
  and the surface velocity field (arrows) on a particular 
  portion of the longitude-latitude plane in case CAt.}
\label{fig_ar_vitesse}
\end{figure*}

Panel 3 confirms the results in the isentropic case that showed that
the poleward drift of the flux tubes due both to the uncompensated
magnetic curvature force and to the Coriolis force acting on the tube
is more active at high latitudes. Indeed, it is clear that at $60\degr$ and $75\degr$, as soon as the tube has started
rising, it is strongly deviated from the radial trajectory. We recover
the particular behaviour of the tube introduced at $75\degr$ which
undergoes an equatorward drift because of the prograde flow being
created in its interior.
However,
this deviation to the radial trajectory is less pronounced than in the
isentropic cases, where for instance after 6 days of evolution, a tube
initially located at $60\degr$ had deviated by $3.8\degr$. In the same
case here, the deviation angle hardly reaches $2.8\degr$ at the same
time, i.e. $26\%$ less. This difference is mainly due to the longitudinal flow appearing in the tube interior as 
soon as the tube begins to rise, which is much stronger in the isentropic cases than in the convective ones.
This can be understood by considering the non-axisymmetric
  deformation of the tube in the convective case. This leads to
  friction between the magnetic structure and its surroundings which
  in turn transfers angular momentum to the mass elements in the tube
  and therefore leads to less retrograde motion.
For instance, after 4 days of evolution, the tube embedded in the isentropic layer has created a longitudinal flow 
of about $-30\,\rm m.s^{-1}$ whereas the tube embedded in the convective zone has not developed any significant zonal motion, it
is still rotating at the same velocity as its surroundings because the non-uniform rotation in radius plays a role 
in the conservation of the flux rope angular momentum.
As a consequence, the intensity of magnetic field needed for tubes to rise radially may be overestimated in the isentropic case. 
W now move to the study of the influence of the meridional flow, which may also act to advect the magnetic structure away 
from the radial trajectory.

\subsection{Meridional circulation}

As we said in the first section, meridional flows are maintained by
buoyancy forces, Reynolds stresses, pressure gradients, Maxwell
stresses and Coriolis forces acting on the differential
rotation. Since these relatively large forces nearly cancel one
another, this circulation can be thought of as a small departure from
(magneto)geostrophic balance, and the presence of a localised magnetic
field can clearly influence its subtle maintenance.

Indeed, for two different initial
magnetic intensity in the flux rope, it is interesting to focus on the streamfunction of the background
meridional flow. Figure \ref{figure_circmerid} shows the position of
the flux rope close to the end of its evolution through the CZ,
superimposed to the streamfunction of the meridional
velocity. We clearly see that the situation is different in the
two cases. In the strong B case, the contours of the MC streamfunction
are concentrated around the flux rope and are very symmetric with
respect to the tube apex. In this case, this observed velocity field
is created by the flux rope itself through the back-reaction
of the Lorentz force on the flow and it is completely dominant
compared to the background velocity. This velocity field structure,
characterised by a strong upflow at the tube apex and two downdrafts
at each side of the tube drives the tube radially upward, without any
latitudinal drift since the magnetic structure is not sensitive to the
background MC.  The situation is clearly different in the weak B
case. In this case, the velocity field created by the tube itself is
of the same order as the background meridional flow and thus the rope
is very likely to be advected in a particular direction whether it is
embedded in a poleward or in an equatorward flow. Here we note that by
the time it reaches the top of the domain, the tube is drifting
northward partly because of the poleward drift phenomenon we mentioned
in the isentropic case. Thus the tube ends up in a poleward flow at
this particular longitude, which reinforces the poleward advection of
the magnetic structure.

\begin{figure*}
	\centering
	\includegraphics[width=17cm]{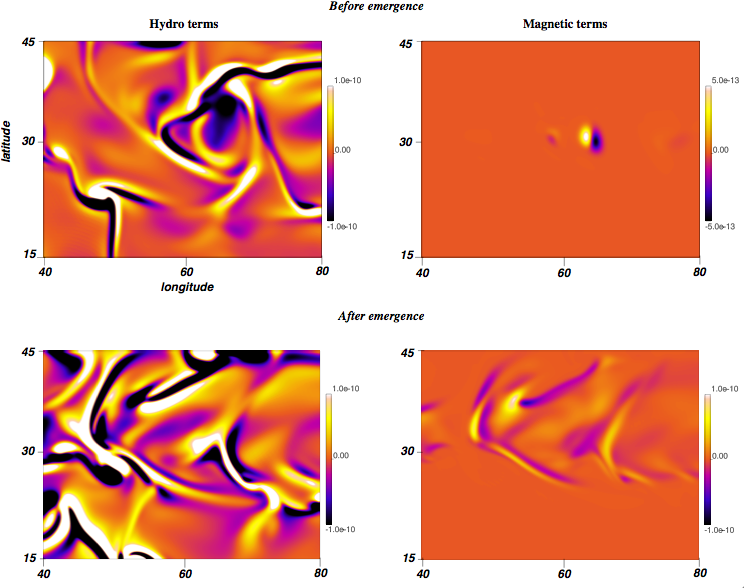}
	\caption{Hydrodynamical and magnetic terms playing a role in
	the evolution equation of the radial vorticity, just before
	emergence and after emergence.} 
\label{figure_radvort}
\end{figure*}

Several observational studies with MDI/SOHO data \citep{Haber03, Haber04, Hindman, Gizon1, Gizon2, Svanda} showed that emergence of new magnetic flux could
generate perturbations on the observed surface horizontal
flow. Consequently, we can focus our study on the modification of this
horizontal flow by the emergence of our flux tube modulated by
convection. Fig. \ref{fig_ar_vitesse} shows the superimposition of the
emerging radial magnetic field and the horizontal velocity field close
to the upper limit of the domain ($0.96 R_\odot$) for case CAt during
the emergence process, and until the intensity of the emerging radial
field has reached a value of about $1500 \, \rm G$. On the first
panel, the magnetic flux has hardly emerged (the intensity of the
radial field is below $10^{-2} \, \rm G$ and the horizontal velocity
field thus presents a pattern which is almost not modified by the
magnetic tube. We then get the emergence of the magnetic structure,
which is showed by the growing intensity of the radial field. 
As the structure emerges, the changes in the horizontal velocity field
due to the magnetic forces are slight but visible. Indeed, the
intensity of the flow is growing because of the presence of magnetic
field, which is especially clear on panels e) and f) around the
positive polarity which is dominant for this particular bipolar patch.
 Moreover, regions of converging flows become more confined between
 the actives longitudes, as we can see for example on panels d), e) and
 f) where the converging flow (associated with a strong downflow
 lane around $70{\degr}$) are particularly concentrated between the
 different emerging bipolar structures.  
Another striking point is the acceleration of the retrograde zonal
flow during emergence, as a result of the azimuthal velocity created within the tube because of angular momentum
conservation.

 Consequently, slight modifications can be seen on the structure of
 the horizontal flow as the magnetic structures emerge. To be more
 quantitative, we focus on the creation of radial vorticity due both
 to hydrodynamical terms and to magnetic terms just before and after
 emergence and how the magnetic field plays a role in this balance to
 locally modify the flow structure. Indeed, following the evolution of
 the radial vorticity enables to track the evolution of the horizontal
 flow profile since the radial vorticity can be expressed as follows:

\begin{equation}
\omega_r=\frac{1}{r\sin\theta} \left[\frac{\partial (\sin\theta v_\phi)}{\partial
\theta}- \frac{\partial v_\theta}{\partial \phi} \right]
\end{equation}

The evolution equation for $w_r$ can be decomposed on 4 hydrodynamical
terms (not depending on $B$ or any of its derivatives) and 3 magnetic
terms, as follows:

\begin{eqnarray}
\frac{\partial \omega_r}{\partial t}&=&\underbrace{\left[ ({\bf \omega_a} \cdot {\bf \nabla}) {\bf
    v} - ({\bf v}\cdot {\bf \nabla}) {\bf \omega_a} 
    -({\bf \nabla} \cdot {\bf v}){\bf \omega_a}  -
    {\bf \nabla} \times (\frac{1}{\bar{\rho}} {\bf \nabla} \cdot {\bf \cal
    D})\right]_r}_{\mbox{hydro terms}}\nonumber \\ 
&+&\underbrace{\left[
    \frac{1}{\bar{\rho} c} \left( ({\bf B} \cdot {\bf \nabla}) {\bf
    j} - ({\bf j} \cdot {\bf \nabla}) {\bf B} -{\bf j} \cdot {\bf
    \nabla} (\frac{1}{\bar{\rho}})\right) \right]_r}_{\mbox{magnetic
    terms}}
\label{eq_vort}  
\end{eqnarray}
 
\noindent with ${\bf \omega_a}$ the absolute vorticity defined by the
relation ${\bf \omega_a}={\bf \nabla \times v} + 2 {\bf \Omega_0}$.

Figure \ref{figure_radvort} shows the value of the sum of only the
hydrodynamical terms and of only the magnetic terms in the radial
vorticity evolution equation \ref{eq_vort}, just before emergence (corresponding to
panel a) of Fig. \ref{fig_ar_vitesse}) and significantly after (corresponding to
panel f) of Fig. \ref{fig_ar_vitesse}).
We note that at the very beginning of emergence, when a small bipolar
patch begins to emerge with the North-South orientation, the magnetic
terms are more than 2 orders of magnitude smaller than the hydro
terms. The latter completely determine the behaviour of the horizontal
flow, especially the advection term (second in the RHS of
Eq. \ref{eq_vort}), which is dominant and peaks at about $10^{-9} \,
\rm s^{-2}$, whereas the dominant magnetic term hardly reaches
$10^{-11} \, \rm s^{-2}$.
On the other hand, when the structure has sufficiently emerged (the
structure and strength of the radial field at this time are showed on
the last panel of Fig. \ref{fig_ar_vitesse}), the magnetic terms
start to play a role in the vorticity generation and thus on the
horizontal flow structure. They have increased by about 2 orders of
magnitude and since the norm of the hydrodynamical terms stays close to
the same values, all those terms begin to equally
compete. We note that the magnetic source terms for radial vorticity
concentrate everywhere the magnetic field gradients are sharp. This
can be seen
especially around the strong positive polarity around 55 degrees of
longitude and 40 degrees of latitude. In this region, the structure
has emerged and thus a strong gradient in longitude of all components
of the the field will act to produce currents which will in turn play
a role in the radial vorticity generation. We thus conclude that the
horizontal flow is modified by magnetic fields preferentially where
strong gradients of field exist, for instance at the edge of the
emerging structure, in agreement with what was concluded from the
analysis of Fig. \ref{fig_ar_vitesse}.

\section{Dynamical evolution in a fully convective shell: influence of the parameters}
\label{sect_param}

We now turn to investigate how the significant parameters of the
reference case influence the flux tube in its rise through the
convection zone in the case where we have a fully turbulent convection
developed in the bulk of the computational domain. We will also look
for new key parameters which may constrain the behaviour of the
magnetic rope during its dynamical evolution.

\subsection{Role of twist}

We saw that the twist of the field lines plays a fundamental role in
the ability of the flux tube to rise cohesively in a stratified layer.
Moreover, observations of active regions show that a certain amount of
twist of the field lines is often detected \citep{Schmieder}, especially in regions called sigmoids. These
sigmoids take the shape of a reversed S in the Northern hemisphere and
of a S in the Southern hemisphere, sign of the hemispheric law for
helicity (which is directly related to the twist) which is found to be
preferentially negative in the North and positive in the South. These
regions are of particular interest because they are known to be
favoured places for the triggering of CMEs or other violent events at
the solar surface. Recent numerical simulations \citep{Torok,Fan3,Amari} show that a twisted flux rope is always present at a certain point in the flux emergence (prior to the emergence for T{\"o}r{\"o}k \& Kliem and built during the emergence for Amari et al.) and that the twist is sometimes the determining factor for the eruption to occur (via the kink-instability for example).

\begin{figure*}
	\centering \includegraphics[width=16cm]{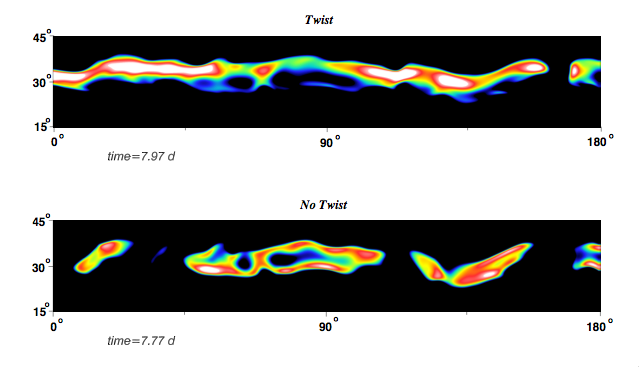}
	\caption{Cut of $B_{\phi}$ at a specific radius after
	approximately 10 days of evolution in case CAt
	(upper panel) and in case CAnt (lower panel). We clearly
	note the splitting of the flux tube in the untwisted case.}
	\label{figure_split}
\end{figure*}

Consequently, both simulations and observations show the fundamental role of the twist of the field lines while flux emerges and a further investigation of this parameter is then particularly important.  
 Figure \ref{figure_split} shows the behaviour of a flux tube
embedded in a fully convective shell in an untwisted case (lower panel)
and a twisted case (upper panel). We recover the fact that a
sufficient twist of the field lines is needed for the tube to maintain
its integrity while it rises through the CZ. Indeed, we note that in
the untwisted case, the tube splits into two parts while it rises
because of the uncompensated vorticity generation created inside the
flux rope by the gravitational torque, as was discussed in
 Sect. \ref{sect_isen} and in Jouve \& Brun \cite{Jouve} in the
reference case. On the contrary, the right panel of
Fig. \ref{figure_split} illustrates the fact that the twisted magnetic structure
has kept its tube-like shape by the time it has almost reached the
top of the shell.

We also note that the deformation of the rope due to the convective up
and down flows is more pronounced in the untwisted case. Indeed, since
the tube splits into two separate concentrations of flux, the two
resulting structures are magnetically less strong and are thus more
sensitive to the surrounding convective motions. Moreover, when the
tube is twisted, magnetic tension acts to prevent convective
downdrafts from penetrating into the magnetic structure. The tube is
thus more cohesive and thus less distorted than in the untwisted case
(even if the modulation in longitude is already very significant) and
is able to reach the top of the computational domain and emerge.

\begin{figure*}
	\centering 
	\includegraphics[width=16cm]{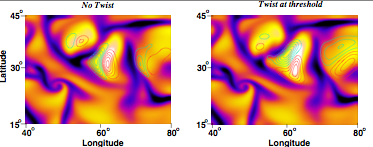}
	\caption{Zoom, seen from above at $r=0.93R_{\odot}$ of an 
	  emerging region in case CAnt (left panel) and a twisted case 
	  where the twist is just above the threshold (right panel). 
	  These snapshots correspond to approximately the same time of 
	  evolution as panel 3 of Fig. \ref{figure_shsl}, we clearly
	  note the difference in the orientation of the main bipolar region.}
	\label{figure_orient}
\end{figure*}

\begin{figure*}
	\centering 
	\includegraphics[width=16cm]{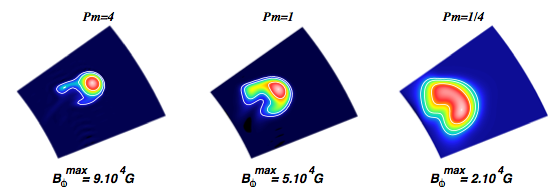}
	\caption{Longitudinal magnetic field after 5 days of evolution
	with $Pm=4$, 1 and 1/4.}
	\label{figure_pm}
\end{figure*}

Figure \ref{figure_orient} shows that the twist of the field lines is
of major interest for the orientation of the emerging bipolar structures as we already saw in the preceding section.
In the non-twisted case, when the tube is sufficiently strong to reach the top of the
CZ, the emerging radial field creates bipolar regions which have the
right East-West orientation. However, these active regions have a
very significant extension in latitude because of the two counter
vortex rolls which drift apart horizontally and this is not what is
observed in the Sun where active regions are very localised in
latitude. If the twist of the field lines just reaches the threshold,
we see that the orientation of the patches becomes East-West quite early
in the emerging process. Indeed in this case, we observe the radial
field coming from the two feet of the arched (because of convective
down and up flows which deform the tube) portion of the tube sooner
than in the very twisted case where the radial field due to the twist
dominates. As a consequence, if we follow the evolution of the tilt
angle for this case as we did for case CAt on Fig. \ref{figure_shslflux},
we see that it becomes East-West much more rapidly after emergence
and above all that the final tilt angle we get is about $-15\degr$,
i.e. closer to the observations at this particular latitude. Moreover,
this case has an initial number of turns of 14 (corresponding to a
pitch angle of about $20\degr$) and thus if we consider that the
emerging region occupies about $20\degr$ in longitude when it has
expanded at the surface, the number of turns in this
particular bipolar region would be of about 0.78, in agreement with the
typical value observed in most active regions.
 This case thus seems to be able to reproduce several interesting features of
active regions such as their orientation (even if the tilt angle is
still high in comparison to observations but could be reduced with
more arched structures), their amount of twist and the field strength
inside the regions of opposite polarities, which is of the order of 1
kiloGauss, as in case CAt.


\subsection{Influence of the diffusivities}
\label{sect_dif}

\begin{figure}
	\centering 
	\includegraphics[width=7.2cm]{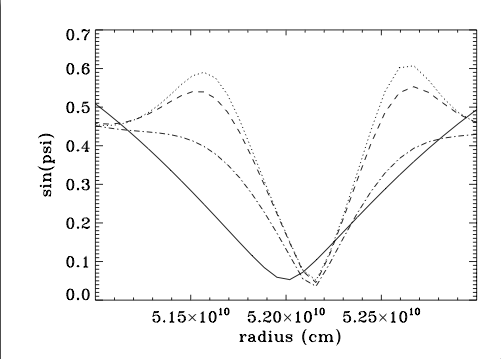}
	\includegraphics[width=7.2cm]{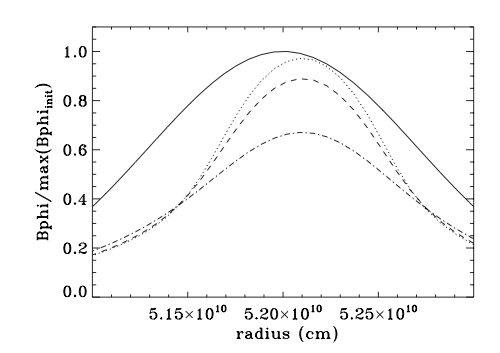}
	\includegraphics[width=7.2cm]{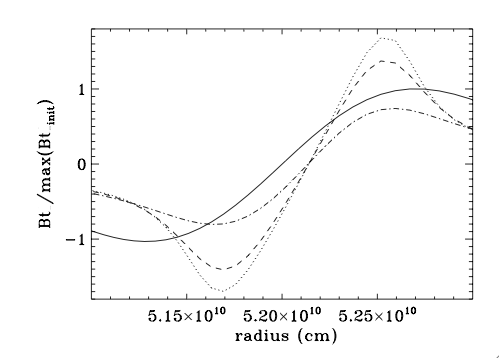}
	\caption{Measure of the sine of the pitch angle (1st panel), of the longitudinal field 
(2nd panel) and of the transverse field (3rd panel) at the initial time (plain line) and after
only 5 hours of evolution for tubes with $Pm=4$ (dotted line), $Pm=1$ (dashed line) 
and $Pm=1/4$ (dash-dotted line).}
	\label{figure_twpm}
\end{figure}

In this section, we investigate the effects of varying the magnetic
diffusivities in our models of flux tubes evolution, keeping all the
other parameters constant. We vary $\eta$ from a value of $1.13 \times 10^{12} \rm cm^2 \,
s^{-1}$ in the
middle of the convection zone (corresponding to a magnetic Prandtl
number of unity) to a value of $2.83 \times 10^{11} \rm cm^2 \,
s^{-1}$ (leading to $Pm=4$) and to a value
of $4.54 \times 10^{12} \rm cm^2 \,
s^{-1}$ (leading to $Pm=1/4$). It has been shown in previous thin
  flux-tube studies \citep{Moreno95} that a strong
  entropy gradient could be built between the tube interior and
  its surroundings during its rise through the CZ. As a consequence of
  higher entropy within the tube, the external gas pressure decreases
  faster than the internal pressure and may finally reach the same
  value, forcing the magnetic pressure to approach zero. The tube apex
  then experiences a so-called explosion which causes this part of the
  tube to stop rising and leads to an amplification of the magnetic
  field in the non-exploded parts \citep[see][for a full MHD treatment
  of this process]{Rempel01}. In our simulations
  where the high diffusion of entropy may wash out the gradients
  responsible for such effects, our tubes do not undergo any explosion
  and stay magnetically buoyant from the base of the CZ to the top of
  our computational domain. However, for this section, we wanted to keep the
  same convective background and at the same time keep the value of $Pr=0.25$
  unchanged since it has proved to be favorable
  to a solar-like differential rotation (Brun \& Toomre 2002; Miesch et
  al. 2006). This has dictated our choice of $\nu$ and $\kappa$ and
  thus we did not consider those parameters as free anymore. At
  the present time, the ASH code uses effective eddy diffusivities to
  represent momentum, heat and magnetic field transport by motions
  which are not resolved by the simulation. They are allowed to vary
  in radius but are independent of
  latitude, longitude and time. This type of treatment for the
  unresolved motions thus affects all spatial scales and it has to be
  stated that this may have a significant influence on the evolution
  of spatially localised structures such as the magnetic flux tubes
  introduced in our simulations and the strong subsequent currents
  created. Nevertheless, we note that the diffusion term as a whole preferentially
  acts where the magnetic field gradients (or equivalently the
  currents) are the strongest. An improved treatment of
  sub-grid-scale motions in ASH is currently being considered, which would
  take into account a spatial dependence of the transport
  coefficients. The influence of this new treatment on our results will
  have to be checked but for this work, we focus on the major
  differences which can already be pointed out between cases at various $Pm$,
  thus showing the particular care with which diffusion has to be
  considered in this type of simulations.

After 5 days in the convection zone, the tubes introduced
with various magnetic diffusivities have evolved in a very different
way, as shown on Fig. \ref{figure_pm}. 
We clearly see the difference in the expansion of the magnetic field concentration as the tube rises,
each cut in the (r,$\theta$) having the same dimensions. Since the diffusive time goes from 
$a^2/\eta=$ 58 days for $Pm=4$ to 14.5 days when $Pm=1$ to 3.6 days when $Pm=1/4$ (using the 
value of the magnetic diffusivity at the base of the convection zone), it is 
straightforward to note that the intensity of the magnetic field retained in the flux rope
after the same time of evolution in the convection zone is strongly decreased (by a factor
4.5 between $Pm=4$ and $Pm=1/4$) when the magnetic 
diffusivity is increased. Moreover, we see that the two sharper and fainter
 structures at each side of the tube
visible on the first panel are completely lost because of the action of diffusion in the 
two other cases.
Finally, the major consequence of
increasing the magnetic diffusivity can be seen on the last panel of
the figure, in the $Pm=1/4$ case. Not only has the tube significantly
expanded compared to the others but it seems to have split apart
because of an insufficient amount of twist to maintain its coherence.
We investigate the evolution of the twist of the field lines in the 3 different cases 
to understand how a high magnetic diffusivity has caused the tube to lose its coherence
during its rise.

Figure \ref{figure_twpm} shows the profile of the sine of the pitch
angle and of the longitudinal and transverse magnetic field at the
location of the flux tube at the starting time and after 5 hours of
evolution for the various cases. We note that the magnetic structures in the $Pm=1$ and
$Pm=4$ cases stiffen in comparison to the initial configuration, a
feature which is mainly due to the creation of current sheets ahead of
the flux tube when it begins its rise through the CZ.

The main results that we deduce from
this analysis is that the transverse field gets a sharper structure
than the longitudinal field as soon as the tube begins its
rise. This
property leads to a faster diffusion of transverse field than of
longitudinal field. Indeed, panel 2 shows that in the $Pm=1/4$ case,
the maximum of $B_{\phi}$ is $69\%$ of the maximum for the $Pm=4$ case
whereas the maximum of the transverse field only reaches $42\%$ of the
maximal transverse field for the less diffusive case. As a consequence, the pitch angle is strongly reduced by this faster diffusion of transverse fields, the tension force is not sufficient anymore to counteract the vorticity generation inside the tube and thus the pitch angle quickly goes under the threshold value needed to maintain the tube coherence.

\subsection{Influence of the tube radius}

We now turn to investigate the influence of the tube radius, which is closely linked to the study of variations of the magnetic diffusivity which we addressed in the preceding section. Indeed, since the diffusive time is proportional to the square of the tube radius, magnetic diffusion will act faster on smaller tubes and we must thus take into account its potential effects on our tubes. Consequently, we choose to compute models with $Pm=4$, which will limit the effects of diffusion, and we modify the tube radius from $10^9 \,\rm cm$ to $2\times 10^9 \,\rm cm$ and to $5\times 10^8 \,\rm cm$. The initial magnetic field is chosen to be equal to $5 B_{eq}$ and the sine of the pitch angle equal to 0.5, as in our standard case CAt. To compute these models, a very high resolution is needed, in particular, in the smallest tube case ($5\times 10^8 \,\rm cm$), we use 1024 points in latitude, 2048 in longitude and 450 in radius, leading to a number of points to describe the tube section of 26 points in radius and 10 points in latitude. In the $10^9 \,\rm cm$ case with $Pm=4$, we also increase the resolution in latitude compared to our previous cases (1024 points instead of 512, although the results are qualitatively similar) but keep 256 points in radius, we thus end up with $N_r=32 \times N_{\theta}=20$ points to resolve the tube section. Finally, for the $2\times 10^9 \,\rm cm$ tube, the number of points in the tube is also $N_r=32 \times N_{\theta}=20$ (as the total resolution in latitude is now 512). In all cases, the number of points is thus significant enough to have a good description of the magnetic field profile inside the rope.

\begin{figure}
	\centering 
	\includegraphics[width=9.8cm]{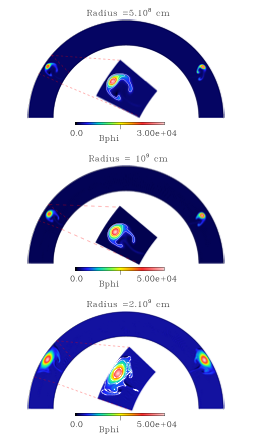}
	\caption{Cut of $B_{\phi}$ in the Northern hemisphere at a
	  specific 
	  longitude for tubes with an initial radius 
	  of $5\times 10^8 \,\rm cm$ (top panel), $10^9 \,\rm cm$ (mid-panel)
	  and 
	  $2\times 10^9 \,\rm cm$ (bottom panel) after about 6.6 
	  days of evolution in a convective model with $Pm=4$.}
	\label{figure_radius}
\end{figure}

Figure \ref{figure_radius} shows the result of the study of the influence of the tube radius on its evolution in the convection zone. We focus on the structure of the longitudinal field $B_{\phi}$ at two different longitudes and at a specific time, close to the end of the evolution (after 6.6 days).
Not surprisingly, the concentration of magnetic flux in the tube interior is broader when the tube is larger and the amount of flux retained in the tube is smaller in the $5\times 10^8 \,\rm cm$-radius tube since the magnetic diffusion has started to play a significant role, even if choosing $Pm$ to be equal to 4 made the diffusive time to be about 14 days in the smallest tube case.
As a consequence, we note on the first panel that the convection has acted to modulate the tube in longitude since the left structure has evolved very differently from the right one. This asymmetry is also visible on the second panel (where the tube radius was originally $10^9 \,\rm cm$) but is less clear on the largest tube calculation, for which the competition with convective motions is in favour of the magnetic structure. 
The main differences in the evolution of these tubes with various
radii reside in the wake that they create during their rise. We can
focus on these differences by zooming on the section of the tube at a
particular longitude, as shown on Fig. \ref{figure_radius}. In the
smallest tube cases, we clearly see two sharp structures being created
at both sides of the magnetic rope and one central tail (especially
visible on the 2 last panels) which enables to approximately follow
the trajectory of the tube axis as it rose through the CZ. These
structures and their properties were studied in great detail in Emonet
\& Moreno-Insertis (1998) and their analysis apply to what we
obtain in our simulations. We can add to this study that a
modification of the tube extension in radius leads to a different
evolution of the wake and thus of the vorticity distribution inside
the tube. Indeed, the wake extends further behind the tube apex when
the tube is smaller, the 2 sidelobes where the vorticity is
concentrated are located significantly behind the apex whereas in the
bigger radius case, the 2 sidelobes appear to stay very close to the
main flux concentration, as was seen in previous calculations in
Cartesian geometry.  

\section{Conclusion}

One of the main goals of this work was to investigate what type of
emerging structures we obtain in the upper part of the convection zone
when we introduce an axisymmetric flux tube at its base. We saw that
all the various flows existing in the convection zone could strongly
influence the behaviour of the tube while it rises. If the tube is
sufficiently weak to be sensitive to the presence of mean flows and
turbulent convection close to the surface, we saw that an azimuthal
modulation was created by convective motions. This modulation in
longitude on the magnetic structure produces arched regions, the
center of which will emerge before the ``sides'' (or the ``feet''). As
a consequence, in the first phases of emergence, only a portion of the
tube is visible at the surface and emerges as a bipolar region. The
orientation of such a bipolar structure will first be North-South but as
the emergence proceeds, different processes will act to produce the
tilt angle corresponding to the statistical Joy's law. As was pointed
out before, both the Coriolis force acting differently on the two legs
of the loop and the twist of the field lines are able to produce an
angle compared to the East-West direction before and during
emergence \citep[e.g.][]{Fan4}. 
We here showed an example of a flux tube possessing an amount of
twist just above the threshold. This simulation can reproduce several
characteristics of active regions, namely the amount of twist in the
bipolar structure, the magnetic field strength in each polarity and the
orientation of the two polarities. 
We moreover emphasise that convective motions advecting separately the two
opposite polarities of the patch of magnetic field can also be a source of the tilting of
active regions, which could not have been investigated in previous Cartesian
or non-convective studies. To disentangle between those various
physical processes acting to produce the tilt angle necessary for some
kind of dynamos to work such as Babcock-Leighton dynamos \citep[see][]{Dikpati99}, we now
need to concentrate on an individual active region emerging at the
solar surface and its particular morphological and dynamical
properties. To do so, we plan to investigate the rise of non-uniformly
buoyant flux tubes from the solar interior to the surface
in a fully convective environment possessing mean flows, as in the
present work. 
 
Indeed, we showed in this study the particular effect of differential rotation
on tubes introduced at different latitudes. As rotation has the
property to slow tubes down during their rise and that rotation is
stronger at lower latitudes, it would imply that
tubes emerging at lower latitudes would have to be more intense to make
their way up to the surface.
The mean meridional flow proves to have smaller effects on the flux
tube rise but may well modify the trajectory of structures slightly in
superequipartition with the strongest downflows at the base of the
convection zone. Moreover, our simulations show that the magnetic
terms can play a significant role in the horizontal flow maintenance close to the
surface and thus that the appearance of magnetic patches at the top of
our domain locally modifies the
surface flow structure.
Mean flows should thus be taken into account in future simulations of
rising magnetic structures, since their interactions with flux ropes are far from
being negligible. 

We now need to consider the introduction of such flux tubes in
a magnetised environment where different scales would interact and the
dynamo field would probably modify the results of the present
study. In particular, reconnection in the interior of our
computational domain between our well-defined flux tube and a more
turbulent chaotic small-scale field is likely to modify the amount of
twist of the field lines contained in the rope. Indeed, how twist is
created in the solar interior and how it is modified during the rise
of magnetic structures are still questions to be addressed. Several
observational studies of helicity in active regions \citep{Schmieder} tend to show that
a systematic twist of the field lines can be observed but the
intensity of which would be small compared to what is needed in
simulations for tubes to rise coherently from the base of the CZ to
the surface. We thus need to reconcile the theoretical and observational
approaches in studying the evolution of the twist of the field lines
of a flux tube embedded
in a realistic magnetised convection zone. We plan to do so in a
future work.
Moreover, to allow some direct comparison to observations, the
  implementation of a stable layer in the ASH code in which a full MHD treatment of the emergence will be
  applied is currently worked on. The results of this more
  realistic upper boundary and the emergence at the top of this new
  domain will be the topic of a following paper.

\acknowledgments
We wish to thank the organisers of the Flux Emergence Workshops held in
St Andrews in June 2007 and in Kyoto in October 2008 and of the KITP dynamo
program in Santa Barbara where a significant part of this work was done. We also thank
Nic Brummell, Brigitte Schmieder, Guillaume Aulanier and Yuhong Fan
 for their very helpful and fruitful comments and the referee
   Manfred Sch{\"u}ssler for his very constructive report which lead
   this article to be significantly improved. We thank PNST for
   partial funding through the work group ``Interfaces physiques et
   couplage de codes'' and LJ acknowledges support
   by STFC. The authors also wish to acknowledge funding by the
   European Research Council through grant ERC-STG STARS2 (www.stars2.eu).

\end{document}